# Role of Native and Zwitterionic Glycine in Electron Attachment to DNA: From Dipole-Bound to Solvent-Bound Doorway States

Ankita Gogoi[1], Jishnu Narayanan S J[1], Sujan Mandal[1] and Achintya Kumar Dutta[1,*]

[1]*Department of Chemistry, Indian Institute of Technology Bombay, Powai, Mumbai-400076, India*

## Abstract

Electron attachment to DNA is strongly influenced by its molecular environment, yet the role of amino acids under physiologically relevant conditions remains poorly understood. Here, we investigate the effect of native and zwitterionic glycine on electron attachment to thymine using high-level electron-affinity calculations and QM/MM molecular dynamics simulations. Under micro-solvated conditions, electron attachment occurs through a dipole-bound doorway state that evolves into a valence-bound anion via nonadiabatic coupling. The zwitterionic form of glycine strengthens stabilization of the diffuse electron owing to its larger internal charge separation, whereas the stability of the valence-bound anion is determined by the hydrogen-bonding geometry. Barrier-free proton transfer is observed only for specific binding motifs and substantially stabilizes the thymine-centered anion. In bulk solution, the doorway mechanism persists, with a solvent-bound state replacing the dipole-bound state as the initial electron-trapping state. The larger electrostatic field of zwitterionic glycine delays electron localization on thymine, while permanent proton transfer is observed only in selected native glycine trajectories and is absent throughout the present simulations of zwitterionic glycine. Despite these differences in electron-transfer dynamics, both amino acid forms provide similar stabilization of the thymine-centered anion after solvent reorganization. Our results establish the solvent-bound state as the condensed-phase analogue of the dipole-bound doorway state and reveal how amino acid environments modulate electron attachment pathways in realistic DNA systems.

*achintya@chem.iitb.ac.in

# 1. Introduction

The stability of DNA is essential for the survival of living organisms, as damage to the genetic material can lead to mutations, genomic instability, and disease. Such damage includes strand breaks, DNA-DNA and DNA-protein cross-links, and base modifications or excision. DNA damage arises from endogenous sources, such as reactive oxygen species generated during cellular metabolism, as well as exogenous agents, particularly ionizing radiation. Although cells possess efficient DNA repair pathways[1] that maintain genomic integrity, these mechanisms are often insufficient to completely repair severe radiation-induced damage.[2–4] Following Muller's discovery of radiation-induced mutations in 1927[5], considerable effort has been devoted to understanding the molecular mechanisms underlying the biological effects of ionizing radiation. DNA damage induced by ionizing radiation is commonly classified into direct, quasi-direct, and indirect pathways.[2,6] Direct effects occur when ionizing radiation interacts with DNA and may lead to ionization or excitation.[2,7–9] Indirect effects result from the interaction of high-energy radiation with surrounding water molecules or proteins. This interaction can either excite or ionize them to form cation radicals and secondary electrons. Both of these species can interact with DNA and induce structural changes.[10,11] Once generated, secondary electrons undergo inelastic collision with the surrounding aqueous environment, producing low-energy electrons (LEEs) with energies ranging from ~0 to 20 eV.[12] As LEEs continue to interact with the surrounding medium, they undergo a sequence of solvation and relaxation processes, eventually becoming thermalized and forming fully solvated electrons ($e^-_{aq}$) within a picosecond timescale. Along this path, they generate potentially harmful intermediates, such as quasi-free electrons ($e^-_{qf}$) followed by pre-hydrated electrons ($e^-_{pre}$), which rapidly undergo solvation within ~500 fs to ultimately yield fully solvated electrons ($e^-_{aq}$).[13–16] Current evidence suggests that quasi-free electrons are primarily responsible for electron-attachment-induced strand breaks, whereas prehydrated and fully solvated electrons are considerably less reactive towards DNA in this regard.[10,17–21]

The initial discovery of LEE-induced DNA damage[22,23] highlighted the crucial role of LEEs in causing radiation-induced damage, prompting extensive experimental[24–32] and theoretical[33–45] investigations to understand how LEEs interact with DNA. These studies emphasized the transient negative ions (TNIs), formed when an incoming free electron is temporarily captured by an unoccupied molecular orbital of a neutral DNA subunit. These transient negative ions correspond to resonance states, which can decay via autodetachment, resulting in the release of an extra electron and reverting to the neutral molecule, or they may undergo dissociative electron attachment (DEA), leading to bond cleavage and the formation of an anion and a neutral radical.[46,47] A detailed understanding of these TNIs provides valuable insights into electron-induced processes that can be utilized to improve the efficacy of radiosensitizers. [48–52] These resonance states lie energetically above the ground state of the neutral molecule and are therefore characterized by negative vertical electron affinity. Electron attachment to gas phase DNA subunits can also lead to the formation of bound anionic states, which have positive electron affinities.[53–56] One of the prominent types of bound states observed in DNA is the dipole-bound state, which arises due to charge-dipole interaction.[57,58] The existence of these dipole-bound states is attributed to the strong dipole moment of DNA. In this case, the excess electron is loosely bound and primarily resides outside the molecular framework.[56] Another type of bound anionic state is the valence-bound one, in which the additional electron density is localized on the molecule and can cause significant alterations in geometry relative to the neutral molecule.[55,59] Previous studies have established that dipole-bound states often serve as doorway states for the formation of valence-bound anions.[59–63] Similar doorway mechanisms have also been identified for condensed-phase nucleobases, where solvent-bound states replace dipole-bound states.[64–68] Given that the surrounding environment significantly affects the dynamics of electron attachment, it is essential to investigate the role of the medium. In the cellular context, DNA is closely associated with proteins, and hence understanding the influence of neighboring amino acids is crucial to perceive the cell response to radiation.

Experiments involving irradiation of DNA with LEEs have revealed that higher amino acid concentrations reduce LEE-induced DNA damage.[69–74] Solomun and Skalicky reported that protein-bound DNA is less susceptible to LEE-induced damage (~3 eV) than isolated DNA.[72] Ptasinska et al. attempted to study the role of proteins by exposing a short single-strand DNA, specifically a tetramer (GCAT), to 1 eV electrons in the presence of amino acids (glycine and arginine).[73] They reported that higher concentrations of amino acids exhibit a protective effect against LEE-induced damage, likely due to van der Waals interactions that shield the DNA molecules. Additionally, gas phase studies on anionic complexes of uracil-glycine, uracil-alanine, and thymine-glycine have revealed that a barrier-free proton transfer can occur from the carboxylic group of glycine to the nucleobase, stabilizing the excess negative charge localized on it.[69–71] First principles molecular dynamics simulations confirmed glycine's protective role via proton transfer or electron scavenging.[75] Amino acids have also been shown to repair guanine radicals.[76–78] DFT studies have revealed that dehydrogenated guanine anions formed via DEA can be retrieved to their normal form through a two-electron-coupled proton transfer (TECPT) mechanism.[77] Recent works by Verma et al.[67] and Wang et al.[74] support amino acids' role in mitigating radiation-induced damage to DNA. Despite these advances, nearly all theoretical studies have focused on the native form of amino acids, whereas amino acids predominantly exist in their zwitterionic form under physiological conditions.[79] Moreover, most available studies are limited to micro-solvated clusters, leaving it unclear whether the mechanisms identified under such conditions remain valid in bulk amino acid environments. Sarma et al. reported that in an arginine-guanosine dimer complex, the electron attaches more readily to the guanine moiety and the DEA process occurs efficiently than in guanosine alone. They also noted, however, that this effect may vary at higher arginine concentrations.[80]

Because the excess electron in valence-bound states is primarily localized on the nucleobase, thymine provides a suitable model for investigating the fundamental mechanisms of electron attachment. Glycine was chosen to represent the amino acid environment because it is the smallest amino acid and is present in the zwitterionic form at physiological pH. The aim of this study is to understand the effect of zwitterionic amino acid on the electron attachment pathways to nucleobases by taking thymine and glycine as a model system.

## 2. Computational Details

To investigate the influence of the environment on electron attachment, we considered both micro-solvated and bulk-solvated models. The microsolvation study included thymine-water, thymine-native glycine, and thymine-zwitterionic glycine complexes. Since multiple hydrogen-bonding motifs are possible in the dimer complexes, we performed conformational sampling using the CREST software[81] to generate different conformers. This is particularly important because conformational changes can modify the electronic structure and hydrogen-bonding network, leading to distinct electron attachment behavior. Recent work by Sarma and co-workers has also demonstrated that different conformers of dipeptide molecules exhibit distinct electron attachment behaviors.[82] However, the dimer complex with zwitterionic glycine was found to be unstable, as a proton transferred from the ammonium group to the carboxyl group, reverting the structure to its native form. To retain the zwitterionic state, an additional water molecule was added to the thymine-zwitterionic glycine complex, which could stabilize the structure via H-bonding interactions with the charged moieties of zwitterionic glycine. The fifteen lowest energy conformers obtained from CREST were then optimized at the RI-MP2/def2-TZVP level of theory. These conformers were further classified according to their hydrogen-bonding motifs. The lowest energy conformer from each category was selected for further analysis. The EA-EOM-DLPNO-CCSD[83]/aug-cc-pVTZ level of theory was employed to evaluate the electron affinities of dimer complexes. The NORMALPNO setting was used for the EA-EOM-DLPNO-CCSD[83] calculations. To account for the diffuse nature of the anion wavefunction, 5s, 5p, and 4d diffuse functions were added in an even tempered manner[84] to the atom closest to the positive end of the dipole moment vector. Bulk-solvated thymine-solvent systems were then prepared using a CHARMM-compatible force field for thymine, glycine, and the TIP3P model for water. The topology and parameters for native and zwitterionic glycine were obtained from CHARMM generalized force field (CGenFF) and CHARMM36 MM force field, respectively. Simulation boxes (40 Å) were constructed using experimental densities of water and glycine at room temperature. Subsequently, three distinct boxes were set up consisting of thymine, 2657 water molecules, 827 native glycine, and 827 zwitterionic glycine molecules respectively. The classical molecular dynamics (MD) simulations were performed with NAMD 2.14.[85] The non-bonded cutoff distance was set to 12 Å. Each system was subjected to energy minimization and subsequently heated to 300 K with positional constraints on thymine. The thymine-water, thymine-native glycine and thymine-zwitterionic glycine systems were then equilibrated for 1 ns, 20 ns and 10 ns, respectively, under conditions of the NVT ensemble at 300 K with periodic boundary conditions. The particle mesh Ewald (PME) technique was used to account for electrostatic interactions. For the production run, the NPT ensemble was used. A 10 ns production run was performed for thymine-water, 120 ns for thymine-native glycine, and 50 ns for thymine-zwitterionic glycine using a simulation time step of 2 fs. Three independent production simulations were performed for each system to ensure adequate sampling. The configurations extracted from the production run were then used for the quantum mechanics/molecular mechanics (QM/MM) simulations. To account for the surrounding environment, solvent molecules belonging to the first solvation shell were also included in the QM region along with thymine. The number of solvent molecules to be included in the first solvation shell was determined using cutoff distances corresponding to the first minima of the radial distribution functions (RDFs) obtained from classical MD simulations. The RDFs were calculated between the hydrogen-bond donor and acceptor atoms of thymine and those of the surrounding molecules (water, native glycine, and zwitterionic glycine). The chosen cut-offs were 2.7 Å for thymine-water, 2.8 Å for thymine-native glycine, and 2.6 Å for thymine-zwitterionic glycine. The molecules in the QM region were treated at the RI-BP86/def2-SVP level of theory and the remaining molecules were treated at the MM level using the same force field used in classical MD simulations. The electron attachment dynamics of all three systems were analyzed using QM/MM dynamics in the presence of an excess electron. For each system, three QM/MM simulations were conducted for 10 ps with a time step of 0.5 fs. Following the simulations, QM/MM single point calculations were carried out on snapshots extracted from the QM/MM trajectories using the EA-EOM-DLPNO-CCSD[83]/aug-cc-pVDZ(+5s5p4d) method with a LOOSEPNO setting, maintaining the same QM/MM partition as employed during the QM/MM dynamics. All QM and QM/MM calculations were carried out using ORCA 5.0.3.[85–87]

## 3. Results and Discussion

### *3.1 Effect of microsolvation*

The thymine-glycine complexes generated from conformational sampling were classified into three distinct categories according to their characteristic hydrogen bonding patterns. These categories were defined based on the specific binding site of glycine on thymine (Figure 1):

1. Glycine interacting with H11 and O7 of thymine.
2. Glycine interacting with H12 and O9 of thymine.
3. Glycine interacting with H12 and O7 of thymine.

This classification facilitates comparison of how the interaction geometry influences the stability of the anionic states. For clarity, we denote the three categories of thymine-native glycine conformers as Thy-NGly1, Thy-NGly2, and Thy-NGly3, respectively. Similarly, the corresponding thymine-zwitterionic glycine complexes are labelled as Thy-ZGly1, Thy-ZGly2, and Thy-ZGly3. The optimized structures of Thymine-Native Glycine and Thymine-Zwitterionic Glycine dimer complexes are shown in Figures 2 and 3, respectively. The left-hand side of both figures contains the neutral structure, and the right-hand side contains the anionic structure. In both Thy-NGly2 and Thy-ZGly2, the carbonyl oxygen (O9) of thymine accepts a proton from glycine in their equilibrium anion geometry. This behavior reflects the localization of the excess electron on the O9-side π orbital of thymine, which increases the basicity of this site. This proton transfer plays a significant role in the stabilization of the anionic state, as will be discussed later. In the remaining two categories, a hydrogen bond exists between the acidic proton of glycine and thymine O7 atom. Although a proton transfer is not observed here, this hydrogen bond length decreases upon going from neutral to anionic geometry.

Further calculations were carried out to determine vertical electron affinity (VEA), vertical detachment energy (VDE), and adiabatic electron affinity (AEA) for all conformers which are defined as

VEA = $E_{neutral}$ (at neutral geometry) – $E_{anion}$ (at neutral geometry)

VDE = $E_{neutral}$ (at anionic geometry) – $E_{anion}$ (at anionic geometry)

AEA = $E_{neutral}$ (at neutral geometry) – $E_{anion}$ (at anionic geometry).

The attachment of an electron can lead to the formation of a dipole-bound anion, where the additional electron density is localized away from the geometric framework. As a result, the nuclear geometry of a dipole-bound anion experiences only minimal distortions compared to its neutral precursor. Consequently, the neutral equilibrium geometry provides a good approximation to the geometry of the dipole-bound state. Because structural relaxation is minimal, the dipole-bound anions have nearly identical VEA and AEA.[88] As shown in Figure 4, the excess electron density in dipole-bound anions, for thymine and monohydrated thymine, is located away from the nuclear framework. The presence of glycine does not alter the qualitative nature of the initial electron attachment, which remains dipole-bound despite changes in the electrostatic environment. The valence-bound anionic states, on the other hand, have the excess electron localized within the nuclear framework, leading to significant geometric distortions, and thus, valence-bound anions are only stable at the anionic geometry. Figures 5 and 6 show the valence-bound anions of all thymine-glycine complexes under consideration.

Table 1 lists the dipole moment (calculated at the Hartree-Fock level of theory), VEA, VDE, and AEA for thymine and thymine-solvent complexes with water and amino acids (calculated at the EA-EOM-DLPNO-CCSD/aug-cc-pVTZ(+5s5p4d) level of theory). The thymine-water complex exhibits an almost identical VEA compared to isolated thymine, while thymine-glycine complexes show higher VEA values than isolated thymine. The increased VEA reflects stabilization of the diffuse dipole-bound electron by the electrostatic field generated by glycine. On the other hand, Thy-NGly1 is very weakly bound vertically with a VEA of 4 meV, likely due to its relatively low dipole moment (3.18 D). The zwitterionic glycine complex shows higher dipole moment value due to the charge separation. Although dipole-bound states generally become more stable with increasing dipole moment, the correlation is not strictly quantitative because the spatial distribution of the electrostatic potential also influences electron binding.[89] Overall, the zwitterionic complexes exhibit larger VDE values than the native complexes. The AEA value is found to be positive for both the dimer complexes of thymine and glycine where a proton is transferred to thymine from the latter. The thymine-glycine complexes with proton transfer, namely Thy-NGly2 and Thy-ZGly2, also exhibit the highest VDE among all other complexes. These results clearly demonstrate that proton transfer substantially stabilizes the valence-bound anion, as evidenced by the simultaneous increase in both AEA and VDE. Thy-ZGly3 also forms an adiabatically bound anion, although no proton transfer occurs from glycine to thymine. This is presumably due to the additional hydrogen bonding with the water molecule, which is absent in Thy-NGly1. This observation demonstrates that proton transfer is not essential for the formation of a stable valence-bound anion.

Two pathways have been recognized for the protective mechanism of amino acids from LEE-induced damage.[67,75] In the first pathway, amino acids act as an electrostatic shield by directly scavenging the additional electron, thereby preventing its attachment to DNA. The second pathway involves stabilizing the excess negative charge on the nucleobase through favorable interactions or even proton transfer. Figures 5 and 6 illustrate the valence-bound state of all the thymine-glycine complexes, showing that the excess electron is localized on the nucleobase. The initial dipole-bound state localizes the excess electron outside the nucleobase framework, thereby delaying population of the chemically reactive valence-bound state. Once the excess electron occupies the π valence state, transfer to the dissociative σ* state becomes kinetically less favorable.[63] Additionally, a proton is transferred from glycine to thymine on the valence-bound anion for category 2 complexes. Gutowski and co-workers, based on their photoelectron spectroscopic experiments and DFT calculations[71] have proposed that an anionic dimer complex of thymine and glycine undergoes a barrier-free proton transfer when the H atom of COOH is coordinated to the O9 atom of thymine. Whereas no proton transfer occurs in complexes with glycine coordinated to the O7 atom of thymine (Figure 1). This difference arises because the excess electron is not localized on the O7 atom, as indicated by the valence-bound state of thymine shown in Figure 4(a). This is consistent with our results where a proton is transferred to the O9 atom in thymine from both native and zwitterionic glycine, thereby forming an adiabatically bound anion. Moreover, the additional hydrogen-bonding with the surrounding molecules can itself provide additional stabilization of the anion. For example, a positive AEA value is observed for the Thy-ZGly3 complex, even though no explicit proton transfer from glycine to thymine takes place. In this case, the hydrogen-bonding interaction between the O9 atom and the water molecule at the anionic geometry (Figure 3(c)) likely stabilizes the excess electron, thereby enabling the formation of an adiabatically bound anion.

Two limiting mechanisms may describe electron-attachment-induced proton transfer: (i) sequential formation of a valence-bound intermediate followed by proton transfer, and (ii) concerted electron-proton transfer (CEPT).[90] To investigate the proton transfer mechanism in thymine-glycine complexes, we have calculated the corresponding minimum energy pathway (MEP) in the valence-bound anion. If the calculated pathway exhibits a potential energy barrier, it suggests that the proton transfer occurs in a two-step process, where the valence-bound anion acts as an intermediate state. On the other hand, if the pathway is found to be barrierless, it would indicate that both the electron and proton are transferred simultaneously, without the formation of any intermediates, consistent with the CEPT mechanism. To determine the MEP, we performed a Nudged Elastic Band (NEB) calculation at the same level of theory used for optimization. We then recalculated the MEP at the EA-EOM-DLPNO-CCSD/aug-cc-pVTZ(+5s5p4d) level of theory using the intermediate geometries obtained from the NEB calculation. Figure 7 illustrates the MEP between valence-bound anion and the final proton transferred product for complexes of thymine with both native and zwitterionic glycine. It can be seen that the proton transfer is a barrierless process, hence, indicating that it is a CEPT reaction.

Previous studies have shown that electron attachment to thymine nucleobase follows a doorway mechanism.[59] To unravel the impact of microsolvation on dipole-bound to valence-bound transition, we have generated the adiabatic potential energy curve (PEC) of ground and first excited states of anion for isolated thymine and all the dimer complexes considered. This was done by tracing a linear transit from the dipole-bound geometry to the valence-bound geometry. The intermediate geometries along this linear transit were calculated using the following expression:

$$R_{\text{int}} = (1-\lambda)R_{DB} + R_{VB}$$

where $R_{DB}$ refers to the geometric parameter (bond length, bond angle and dihedral angle) of the dipole-bound geometry and $R_{VB}$ represents the corresponding parameters for valence-bound geometry. $\lambda$ varies from 0 to 1 to map the transition between the two states, where λ = 0 corresponds to the dipole-bound geometry, while λ = 1 corresponds to the valence-bound geometry.

It can be seen from the Figures 8, 9 and 10 that the adiabatic potential energy curve for the ground state and the excited state come close to each other, showing an avoided crossing which indicates strong interaction between the electronic and nuclear motion. The nature of the electronic wavefunction changes rapidly in the vicinity of the avoided crossing and the Born–Oppenheimer approximation becomes less accurate because of enhanced nonadiabatic coupling.[91] One can switch to diabatic representation by transforming the adiabatic electronic wavefunctions into diabatic states, which retain their character with a change in nuclear coordinates, and the electron-nuclear coupling is replaced by the electronic coupling between the diabatic states.[60,61] We have chosen the dipole-bound and valence-bound states of the anion as the diabatic basis and calculated the coupling between two states by fitting an avoided crossing model potential defined as follows:

$$\begin{pmatrix} V_1 & W \\ W & V_2 \end{pmatrix}.$$

The off-diagonal term W is assumed to be constant and a fourth-degree polynomial as a function of $\lambda$ has been used to obtain the diagonal terms $V_1$ and $V_2$.

$$V_i = a\lambda^4 + b\lambda^3 + c\lambda^2 + d\lambda + v_i^0$$

The fitted electronic coupling (W) provides a quantitative measure of the electron transfer between dipole-bound and valence-bound states.

We have evaluated the rate constant using Marcus theory[92]

$$k = \frac{2\pi}{\hbar^2} |W|^2 \sqrt{\frac{1}{4\pi k_B T \lambda_R}} e^{-(\lambda_R + \Delta G^0)^2 / 4\lambda_R k_B T}$$

where $\lambda_R$ is the reorganization energy, $W$ is the coupling constant and $\Delta G^0$ is the free energy difference between the valence-bound and dipole-bound states, excluding entropy contribution.

Table 2 lists the coupling constants and rate constants for electron transfer in isolated thymine and all complexes of thymine. The rate constant is found to be the highest for thymine-glycine complexes where a proton is transferred from glycine to thymine, namely Thy-NGly2 and Thy-ZGly2. In these cases, the rate increases by three orders of magnitude relative to isolated thymine. For the other two complexes, where proton transfer does not take place, the rate constant decreases by one order of magnitude in the presence of zwitterionic glycine than that observed in the isolated thymine but increases by one order of magnitude with native glycine. Proton transfer lowers the free-energy difference between the dipole-bound and valence-bound states while simultaneously strengthening electronic coupling, thereby accelerating electron transfer. As the formation of valence-bound anion becomes more favorable, it may compete with alternative DNA damage pathways, thereby suppressing electron attachment-induced bond cleavage.

However, in the case of Thy-ZGly complexes, where proton transfer does not take place, the formation of nucleobase-centered anion is kinetically less favored than isolated thymine. It can be seen that the proton transfer is strongly dependent on the specific binding site of glycine to thymine, and the stability of the resulting anion depends on the interaction geometry. Collectively, these results demonstrate that the protective role of amino acids depends not only on their chemical identity but also on their binding geometry and proton-transfer capability.

### *3.2 Effect of bulk solvation*

Although micro-solvated clusters capture the essential electronic structure of electron attachment, they cannot fully represent the dynamic hydrogen-bonding network and solvent fluctuations present in aqueous solution. To account for these effects, we investigated electron attachment in bulk solution using QM/MM molecular dynamics. Similar to the micro-solvated systems, electron attachment in bulk proceeds through both the ground and excited electronic states of the anion. Figures 11–13 illustrate the time evolution of the EA-EOM-DLPNO-CCSD dominant transition orbitals for representative QM/MM trajectories of thymine in bulk water, native glycine, and zwitterionic glycine.

In the presence of bulk water, the excess electron is initially localized on the surrounding water molecules, forming a solvent-bound ground state, while the nucleobase-bound state appears as an excited state (Figure 11). The diffuse electron is preferentially stabilized by the collective electrostatic field of the water rather than by the $\pi^*$ orbitals of thymine. As the water molecules reorganize, the excess electron gradually localizes on the nucleobase, causing the nucleobase-bound state to become the electronic ground state. The solvent-bound to nucleobase-bound transition occurs within approximately 2.5 fs, after which the solvent-bound state becomes the first excited state. Similar solvent-mediated doorway states have previously been reported for bulk-hydrated uracil and cytosine, confirming that this mechanism is a general feature of electron attachment to nucleobases in aqueous solution.[66,93]

A similar sequence of events is observed in the presence of native and zwitterionic glycine (Figures 12 and 13). In both cases, the excess electron is initially trapped by the surrounding glycine molecules, giving rise to a solvent-bound ground state, while the thymine-centered valence-bound state appears as an excited state. As the simulation proceeds, the excess electron transfers to the nucleobase and the thymine-bound state becomes the electronic ground state. However, the characteristic time scale of this transition depends strongly on the surrounding medium. In the native glycine environment, the electron localizes on thymine after approximately 13 fs, whereas in the zwitterionic glycine environment the transition occurs only after approximately 27 fs. The substantially slower localization observed in the zwitterionic system indicates that charge separation within the amino acid stabilizes the diffuse solvent-bound electron more effectively than the native form, the signature of which was also observed in the microsolvation study. The stronger electrostatic field generated by the zwitterionic species therefore delays electron localization on the nucleobase by competing with thymine for the excess electron. The remaining QM/MM trajectories exhibit the same qualitative behavior (Figure S3), indicating that the delayed electron transfer is a robust characteristic of the zwitterionic environment rather than a trajectory-specific event.

The microsolvation study demonstrated that interactions involving the O9 atom of thymine play a central role in stabilizing the valence-bound anion through proton transfer. A similar phenomenon is observed in the bulk native glycine environment, where proton transfer from a neighboring glycine molecule to the O9 atom of thymine accompanies the formation of the thymine-centered anion. This observation is consistent with earlier QM/MM simulations on cytosine in bulk glycine, which also reported proton transfer from the surrounding amino acid following electron attachment.[67] However, unlike the micro-solvated complexes, the proton-transfer dynamics in bulk solution are highly dependent on the instantaneous solvent configuration and therefore differ among the three independent trajectories.

In the first trajectory, the proton initially attached to the carboxyl group of glycine is transferred to thymine after approximately 70 fs (Figure 14). The proton subsequently returns to glycine near 800 fs before transferring back to thymine at approximately 3 ps, where it remains for the rest of the simulation. In contrast, the second and third trajectories involve proton transfer from the amino group of glycine rather than from the carboxyl group (Figures S4 and S5). Following deprotonation of the amino group, an intramolecular proton transfer from the carboxyl group restores the protonated amino functionality, consistent with the higher proton affinity of the amino group relative to the carboxyl group.[94,95] Unlike the first trajectory, the transferred proton returns to glycine in the remaining trajectories. These observations demonstrate that, although proton transfer is possible in bulk native glycine, the detailed reaction pathway depends sensitively on the evolving hydrogen-bonding network surrounding thymine.

The situation is markedly different in the zwitterionic glycine environment. No proton-transfer event is observed during any of the three independent 10 ps QM/MM trajectories (Figure 15 and Figures S6–S7). The absence of proton transfer suggests that the fluctuating hydrogen-bonding network in bulk zwitterionic glycine does not readily generate the specific donor-acceptor geometry required for barrier-free proton transfer, despite the favorable energetics predicted for optimized micro-solvated clusters. Thus, the proton transfer mechanism identified in isolated complexes is not directly transferable to the condensed phase, where continuous solvent reorganization competes with and suppresses proton migration.

To characterize the stabilization of the excess electron during the dynamics, we monitored the instantaneous average vertical detachment energy (VDE) of the anionic ground state for the three solvent environments (Figure 16). In every system, the VDE exhibits a rapid initial increase, reflecting the transition of the excess electron from the diffuse solvent-bound state to the more strongly bound nucleobase-centered state. Although electron localization occurs within only a few femtoseconds, the VDE converges on a significantly longer time scale, requiring approximately 1 ps in bulk water, 4 ps in native glycine, and 3 ps in zwitterionic glycine. This slower equilibration indicates that, even after the electron localizes on thymine, additional solvent reorganization is required to fully stabilize the resulting valence-bound anion. Owing to their smaller size and faster orientational dynamics, water molecules reorganize more rapidly than glycine molecules, leading to faster convergence of the VDE in aqueous solution. Furthermore, the equilibrium VDE of the thymine-bound anion is significantly larger in water (~4.2 eV) than in either glycine environment (~3.4 eV), indicating stronger electrostatic stabilization of the excess electron in water. This trend is consistent with previous studies on bulk-solvated cytosine.[67]

The first native glycine trajectory illustrates the intimate connection between proton-transfer dynamics and anion stabilization. Between approximately 700 fs and 2 ps, the VDE temporarily converges to a lower value (~2.6 eV), corresponding to the interval during which the transferred proton has returned to glycine. After approximately 3 ps, proton transfer back to thymine is accompanied by a second increase in the VDE, ultimately reaching the converged value characteristic of the protonated thymine anion. In the remaining two trajectories, the proton returns to glycine after the thymine-centered anion has already become fully stabilized. These observations indicate that once solvent reorganization has stabilized the valence-bound anion, the electron remains localized on thymine even if the proton subsequently migrates back to glycine. The enhanced stability of the thymine-centered anion is therefore expected to reduce the likelihood of forming dissociative transient negative ions that initiate DNA strand cleavage.

A comparison of the micro-solvated and bulk-solvated systems reveals a fundamental difference in the electron-attachment mechanism. In isolated micro-solvated complexes, stabilization of the valence-bound anion is strongly influenced by proton transfer and the local hydrogen-bonding geometry. In contrast, bulk solution is dominated by dynamic solvent reorganization, which largely determines both the kinetics of electron localization and the stability of the resulting anion. Although proton transfer is observed in native glycine, both native and zwitterionic glycine ultimately stabilize the thymine-centered anion to a similar extent, as reflected by the converged VDE values. The principal distinction between the two environments lies in the initial electron-transfer dynamics: the larger charge separation in zwitterionic glycine stabilizes the solvent-bound doorway state more effectively, thereby delaying localization of the excess electron on thymine. These findings reveal that the solvent-bound state in condensed phases performs a role analogous to that of the dipole-bound state in micro-solvated clusters.

## 4. Conclusions

In this work, we investigated the influence of native and zwitterionic glycine on the electron attachment process to thymine using a combination of high-level electron-affinity calculations and QM/MM molecular dynamics simulations. The microsolvation study demonstrates that electron attachment proceeds through the well-established doorway mechanism, in which a dipole-bound state evolves into a valence-bound anion through non-adiabatic coupling. The larger charge separation in zwitterionic glycine enhances stabilization of the dipole-bound electron, leading to larger vertical electron affinities, whereas the stability of the valence-bound state depends primarily on the local hydrogen-bonding geometry. Proton transfer from glycine to thymine substantially stabilizes the valence-bound anion and proceeds through a barrier-free concerted electron-proton transfer mechanism for favorable hydrogen-bonding motifs.

The electron-attachment mechanism changes qualitatively in bulk solution. Following electron attachment, the excess electron is initially stabilized in a solvent-bound state before localizing on the nucleobase. Native and zwitterionic glycine both delay this localization relative to water, with the effect being more pronounced for the zwitterionic form because of its stronger internal charge separation. Although proton transfer is observed in selected native glycine trajectories, no permanent proton-transfer occur during the present QM/MM simulations of zwitterionic glycine. Nevertheless, both amino acid forms ultimately provide comparable stabilization of the thymine-centered anion after solvent reorganization. These results demonstrate that, in condensed phases, the solvent-bound state plays the same mechanistic role as the dipole-bound state in micro-solvated systems, thereby extending the doorway mechanism to realistic aqueous environments.

The present study provides molecular-level insight into how the amino acid environment modulates electron attachment to nucleobases. Rather than directly preventing electron attachment, amino acids alter the kinetics of electron localization and the stability of the resulting anion through electrostatic interactions, hydrogen bonding, and proton transfer. These findings improve our understanding of radiation-induced electron attachment in biologically relevant environments and provide a framework for future investigations involving larger DNA fragments, peptides, and explicit protein environments.

## Supporting Information

The optimized geometries of isolated thymine, thymine-water, and thymine-glycine complex geometry, molecular orbital plots for additional trajectories are provided in the supporting information.

## Acknowledgments

The authors gratefully acknowledge financial support from IIT Bombay and the Prime Minister's Research Fellowship. The authors also acknowledge IIT Bombay supercomputing facility and C-DAC supercomputing resources (Param Smriti, Param Brahma, and Param Rudra) for computational time.

## References:

(1) Sancar, A.; Lindsey-Boltz, L. A.; Ünsal-Kaçmaz, K.; Linn, S. Molecular Mechanisms of Mammalian DNA Repair and the DNA Damage Checkpoints. *Annu. Rev. Biochem.* **2004**, *73* (1), 39–85. https://doi.org/10.1146/annurev.biochem.73.011303.073723.

(2) Radiation-Induced Damage in DNA. In *Studies in Physical and Theoretical Chemistry*; Elsevier, 2001; pp 585–622. https://doi.org/10.1016/s0167-6881(01)80023-9.

(3) Sage, E.; Shikazono, N. Radiation-Induced Clustered DNA Lesions: Repair and Mutagenesis. *Free Radical Biology and Medicine* **2017**, *107*, 125–135. https://doi.org/10.1016/j.freeradbiomed.2016.12.008.

(4) Berthel, E.; Ferlazzo, M. L.; Devic, C.; Bourguignon, M.; Foray, N. What Does the History of Research on the Repair of DNA Double-Strand Breaks Tell Us?—A Comprehensive Review of Human Radiosensitivity. *IJMS* **2019**, *20* (21), 5339. https://doi.org/10.3390/ijms20215339.

(5) Muller, H. J. Artificial Transmutation of the Gene. *Science* **1927**, *66* (1699), 84–87. https://doi.org/10.1126/science.66.1699.84.

(6) Narayanan S J, J.; Tripathi, D.; Verma, P.; Adhikary, A.; Dutta, A. K. Secondary Electron Attachment-Induced Radiation Damage to Genetic Materials. *ACS Omega* **2023**, *8* (12), 10669–10689. https://doi.org/10.1021/acsomega.2c06776.

(7) Electron Spin Resonance of Radicals in Irradiated DNA. In *Applications of EPR in Radiation Research*; Springer International Publishing: Cham, 2014; pp 299–352. https://doi.org/10.1007/978-3-319-09216-4_8.

(8) Ma, J.; Denisov, S. A.; Adhikary, A.; Mostafavi, M. Ultrafast Processes Occurring in Radiolysis of Highly Concentrated Solutions of Nucleosides/Tides. *IJMS* **2019**, *20* (19), 4963. https://doi.org/10.3390/ijms20194963.

(9) Chapter 31. Gamma- and Ion-Beam DNA Radiation Damage: Theory and Experiment. In *Chemical Biology*; Royal Society of Chemistry: Cambridge, 2020; pp 426–457. https://doi.org/10.1039/9781839162541-00426.

(10) Kohanoff, J.; McAllister, M.; Tribello, G. A.; Gu, B. Interactions between Low Energy Electrons and DNA: A Perspective from First-Principles Simulations. *J. Phys.: Condens. Matter* **2017**, *29* (38), 383001. https://doi.org/10.1088/1361-648x/aa79e3.

(11) Wishart, J. F.; Madhava Rao, B. S. *Recent Trends in Radiation Chemistry*; World scientific: Singapore, 2010.

(12) Alizadeh, E.; Orlando, T. M.; Sanche, L. Biomolecular Damage Induced by Ionizing Radiation: The Direct and Indirect Effects of Low-Energy Electrons on DNA. *Annu. Rev. Phys. Chem.* **2015**, *66* (1), 379–398. https://doi.org/10.1146/annurev-physchem-040513-103605.

(13) Pimblott, S. M.; LaVerne, J. A. On the Radiation Chemical Kinetics of the Precursor to the Hydrated Electron. *J. Phys. Chem. A* **1998**, *102* (17), 2967–2975. https://doi.org/10.1021/jp980496v.

(14) Paik, D. H.; Lee, I.-R.; Yang, D.-S.; Baskin, J. S.; Zewail, A. H. Electrons in Finite-Sized Water Cavities: Hydration Dynamics Observed in Real Time. *Science* **2004**, *306* (5696), 672–675. https://doi.org/10.1126/science.1102827.

(15) Alizadeh, E.; Sanche, L. Precursors of Solvated Electrons in Radiobiological Physics and Chemistry. *Chem. Rev.* **2012**, *112* (11), 5578–5602. https://doi.org/10.1021/cr300063r.

(16) Kumar, A.; Becker, D.; Adhikary, A.; Sevilla, M. D. Reaction of Electrons with DNA: Radiation Damage to Radiosensitization. *IJMS* **2019**, *20* (16), 3998. https://doi.org/10.3390/ijms20163998.

(17) Nabben, F. J.; Karman, J. P.; Loman, H. Inactivation of Biologically Active DNA by Hydrated Electrons. *International Journal of Radiation Biology and Related Studies in Physics, Chemistry and Medicine* **1982**, *42* (1), 23–30. https://doi.org/10.1080/09553008214550881.

(18) K. Kuipers M. V. M. Lafleur, G. Characterization of DNA Damage Induced by Gamma-Radiationderived Water Radicals, Using DNA Repair Enzymes. *International Journal of Radiation Biology* **1998**, *74* (4), 511–519. https://doi.org/10.1080/095530098141384.

(19) Ma, J.; Kumar, A.; Muroya, Y.; Yamashita, S.; Sakurai, T.; Denisov, S. A.; Sevilla, M. D.; Adhikary, A.; Seki, S.; Mostafavi, M. Observation of Dissociative Quasi-Free Electron Attachment to Nucleoside via Excited Anion Radical in Solution. *Nat Commun* **2019**, *10* (1). https://doi.org/10.1038/s41467-018-08005-z.

(20) Ma, J.; Wang, F.; Denisov, S. A.; Adhikary, A.; Mostafavi, M. Reactivity of Prehydrated Electrons toward Nucleobases and Nucleotides in Aqueous Solution. *Sci. Adv.* **2017**, *3* (12). https://doi.org/10.1126/sciadv.1701669.

(21) Kumar, A.; Sevilla, M. D. Role of Excited States in Low-Energy Electron (LEE) Induced Strand Breaks in DNA Model Systems: Influence of Aqueous Environment. *ChemPhysChem* **2009**, *10* (9–10), 1426–1430. https://doi.org/10.1002/cphc.200900025.

(22) Woldhuis, J.; Verberne, J. B.; Lafleur, M. V. M.; Retèl, J.; Blok, J.; Loman, H. γ-Rays Inactivate ϕX174 DNA in Frozen Anoxic Solutions at −20°C Mainly by Reactions of Dry Electrons. *International Journal of Radiation Biology and Related Studies in Physics, Chemistry and Medicine* **1984**, *46* (4), 329–330. https://doi.org/10.1080/09553008414551501.

(23) Boudaïffa, B.; Cloutier, P.; Hunting, D.; Huels, M. A.; Sanche, L. Resonant Formation of DNA Strand Breaks by Low-Energy (3 to 20 eV) Electrons. *Science* **2000**, *287* (5458), 1658–1660. https://doi.org/10.1126/science.287.5458.1658.

(24) Martin, F.; Burrow, P. D.; Cai, Z.; Cloutier, P.; Hunting, D.; Sanche, L. DNA Strand Breaks Induced by 0–4 eV Electrons: The Role of Shape Resonances. *Phys. Rev. Lett.* **2004**, *93* (6). https://doi.org/10.1103/physrevlett.93.068101.

(25) Huels, M. A.; Boudaïffa, B.; Cloutier, P.; Hunting, D.; Sanche, L. Single, Double, and Multiple Double Strand Breaks Induced in DNA by 3−100 eV Electrons. *J. Am. Chem. Soc.* **2003**, *125* (15), 4467–4477. https://doi.org/10.1021/ja029527x.

(26) Abdoul-Carime, H.; Gohlke, S.; Fischbach, E.; Scheike, J.; Illenberger, E. Thymine Excision from DNA by Subexcitation Electrons. *Chemical Physics Letters* **2004**, *387* (4–6), 267–270. https://doi.org/10.1016/j.cplett.2004.02.022.

(27) Zheng, Y.; Cloutier, P.; Hunting, D. J.; Sanche, L.; Wagner, J. R. Chemical Basis of DNA Sugar−Phosphate Cleavage by Low-Energy Electrons. *J. Am. Chem. Soc.* **2005**, *127* (47), 16592–16598. https://doi.org/10.1021/ja054129q.

(28) Sanche, L. Low Energy Electron-Driven Damage in Biomolecules. *Eur. Phys. J. D* **2005**, *35* (2), 367–390. https://doi.org/10.1140/epjd/e2005-00206-6.

(29) Ptasinska, S.; Denifl, S.; Scheier, P.; Illenberger, E.; Märk, T. D. Bond- and Site-Selective Loss of H Atoms from Nucleobases by Very-Low-Energy Electrons (<3 eV). *Angew Chem Int Ed* **2005**, *44* (42), 6941–6943. https://doi.org/10.1002/anie.200502040.

(30) Ptasińska, S.; Denifl, S.; Gohlke, S.; Scheier, P.; Illenberger, E.; Märk, T. D. Decomposition of Thymidine by Low-Energy Electrons: Implications for the Molecular Mechanisms of Single-Strand Breaks in DNA. *Angew Chem Int Ed* **2006**, *45* (12), 1893–1896. https://doi.org/10.1002/anie.200503930.

(31) Zheng, Y.; Cloutier, P.; Hunting, D. J.; Wagner, J. R.; Sanche, L. Phosphodiester and N-Glycosidic Bond Cleavage in DNA Induced by 4–15 eV Electrons. *The Journal of Chemical Physics* **2006**, *124* (6). https://doi.org/10.1063/1.2166364.

(32) Kočišek, J.; Pysanenko, A.; Fárník, M.; Fedor, J. Microhydration Prevents Fragmentation of Uracil and Thymine by Low-Energy Electrons. *J. Phys. Chem. Lett.* **2016**, *7* (17), 3401–3405. https://doi.org/10.1021/acs.jpclett.6b01601.

(33) Barrios, R.; Skurski, P.; Simons, J. Mechanism for Damage to DNA by Low-Energy Electrons. *J. Phys. Chem. B* **2002**, *106* (33), 7991–7994. https://doi.org/10.1021/jp013861i.

(34) Berdys, J.; Anusiewicz, I.; Skurski, P.; Simons, J. Damage to Model DNA Fragments from Very Low-Energy (<1 eV) Electrons. *J. Am. Chem. Soc.* **2004**, *126* (20), 6441–6447. https://doi.org/10.1021/ja049876m.

(35) Berdys, J.; Skurski, P.; Simons, J. Damage to Model DNA Fragments by 0.25−1.0 eV Electrons Attached to a Thymine Π* Orbital. *J. Phys. Chem. B* **2004**, *108* (18), 5800–5805. https://doi.org/10.1021/jp049728i.

(36) Anusiewicz, I.; Berdys, J.; Sobczyk, M.; Skurski, P.; Simons, J. Effects of Base π-Stacking on Damage to DNA by Low-Energy Electrons. *J. Phys. Chem. A* **2004**, *108* (51), 11381–11387. https://doi.org/10.1021/jp047389n.

(37) Bao, X.; Wang, J.; Gu, J.; Leszczynski, J. DNA Strand Breaks Induced by Near-Zero-Electronvolt Electron Attachment to Pyrimidine Nucleotides. *Proc. Natl. Acad. Sci. U.S.A.* **2006**, *103* (15), 5658–5663. https://doi.org/10.1073/pnas.0510406103.

(38) Gu, J.; Xie, Y.; Schaefer, H. F. Near 0 eV Electrons Attach to Nucleotides. *J. Am. Chem. Soc.* **2006**, *128* (4), 1250–1252. https://doi.org/10.1021/ja055615g.

(39) Gu, J.; Wang, J.; Leszczynski, J. Electron Attachment-Induced DNA Single Strand Breaks: $C_{3'}–O_{3'}$ σ-Bond Breaking of Pyrimidine Nucleotides Predominates. *J. Am. Chem. Soc.* **2006**, *128* (29), 9322–9323. https://doi.org/10.1021/ja063309c.

(40) Simons, J. How Do Low-Energy (0.1−2 eV) Electrons Cause DNA-Strand Breaks? *Acc. Chem. Res.* **2006**, *39* (10), 772–779. https://doi.org/10.1021/ar0680769.

(41) Winstead, C.; McKoy, V. Resonant Interactions of Slow Electrons with DNA Constituents. *Radiation Physics and Chemistry* **2008**, *77* (10–12), 1258–1264. https://doi.org/10.1016/j.radphyschem.2008.05.030.

(42) Bhaskaran, R.; Sarma, M. Low-Energy Electron Interaction with the Phosphate Group in DNA Molecule and the Characteristics of Single-Strand Break Pathways. *J. Phys. Chem. A* **2015**, *119* (40), 10130–10136. https://doi.org/10.1021/acs.jpca.5b08000.

(43) Bhaskaran, R.; Sarma, M. The Role of the Shape Resonance State in Low Energy Electron Induced Single Strand Break in 2′-Deoxycytidine-5′-Monophosphate. *Phys. Chem. Chem. Phys.* **2015**, *17* (23), 15250–15257. https://doi.org/10.1039/C5CP00126A.

(44) Kanazawa, Y.; Ehara, M.; Sommerfeld, T. Low-Lying Π* Resonances of Standard and Rare DNA and RNA Bases Studied by the Projected CAP/SAC–CI Method. *J. Phys. Chem. A* **2016**, *120* (9), 1545–1553. https://doi.org/10.1021/acs.jpca.5b12190.

(45) Fennimore, M. A.; Matsika, S. Electronic Resonances of Nucleobases Using Stabilization Methods. *J. Phys. Chem. A* **2018**, *122* (16), 4048–4057. https://doi.org/10.1021/acs.jpca.8b01523.

(46) Arumainayagam, C. R.; Lee, H.-L.; Nelson, R. B.; Haines, D. R.; Gunawardane, R. P. Low-Energy Electron-Induced Reactions in Condensed Matter. *Surface Science Reports* **2010**, *65* (1), 1–44. https://doi.org/10.1016/j.surfrep.2009.09.001.

(47) Recent Progress in Dissociative Electron Attachment. In *Advances In Atomic, Molecular, and Optical Physics*; Elsevier, 2017; pp 545–657. https://doi.org/10.1016/bs.aamop.2017.02.002.

(48) Zdrowowicz, M.; Chomicz, L.; Żyndul, M.; Wityk, P.; Rak, J.; Wiegand, T. J.; Hanson, C. G.; Adhikary, A.; Sevilla, M. D. 5-Thiocyanato-2′-Deoxyuridine as a Possible Radiosensitizer: Electron-Induced Formation of Uracil-C5-Thiyl Radical and Its Dimerization. *Phys. Chem. Chem. Phys.* **2015**, *17* (26), 16907–16916. https://doi.org/10.1039/C5CP02081F.

(49) Muftakhov, M. V.; Shchukin, P. V.; Khatymov, R. V. Thymidine and Stavudine Molecules in Reactions with Low-Energy Electrons. *Radiation Physics and Chemistry* **2021**, *184*, 109464. https://doi.org/10.1016/j.radphyschem.2021.109464.

(50) Ma, J.; Bahry, T.; Denisov, S. A.; Adhikary, A.; Mostafavi, M. Quasi-Free Electron-Mediated Radiation Sensitization by C5-Halopyrimidines. *J. Phys. Chem. A* **2021**, *125* (36), 7967–7975. https://doi.org/10.1021/acs.jpca.1c05974.

(51) Kumar, S.; Sarma, M. Dissociative Electron Attachment to Halogenated Nucleotides: A Quest for Better Radiosensitizers. *Phys. Chem. Chem. Phys.* **2024**, *26* (39), 25524–25532. https://doi.org/10.1039/D4CP02258K.

(52) Kumar, S.; Sarmah, M. P.; Sarma, M. A Theoretical Study of Electron Attachment to Uracil and 5-Halouracil. *ChemPhysChem* **2026**, *27* (13), e202500908. https://doi.org/10.1002/cphc.202500908.

(53) Hendricks, J. H.; Lyapustina, S. A.; De Clercq, H. L.; Snodgrass, J. T.; Bowen, K. H. Dipole Bound, Nucleic Acid Base Anions Studied via Negative Ion Photoelectron Spectroscopy. *The Journal of Chemical Physics* **1996**, *104* (19), 7788–7791. https://doi.org/10.1063/1.471482.

(54) Desfrançois, C.; Abdoul-Carime, H.; Schermann, J. P. Electron Attachment to Isolated Nucleic Acid Bases. *The Journal of Chemical Physics* **1996**, *104* (19), 7792–7794. https://doi.org/10.1063/1.471484.

(55) Roca-Sanjuán, D.; Merchán, M.; Serrano-Andrés, L.; Rubio, M. *Ab Initio* Determination of the Electron Affinities of DNA and RNA Nucleobases. *The Journal of Chemical Physics* **2008**, *129* (9), 095104. https://doi.org/10.1063/1.2958286.

(56) Tripathi, D.; Dutta, A. K. Bound Anionic States of DNA and RNA Nucleobases: An EOM-CCSD Investigation. *Int J of Quantum Chemistry* **2019**, *119* (9), e25875. https://doi.org/10.1002/qua.25875.

(57) Lykke, K. R.; Mead, R. D.; Lineberger, W. C. Observation of Dipole-Bound States of Negative Ions. *Phys. Rev. Lett.* **1984**, *52* (25), 2221–2224. https://doi.org/10.1103/PhysRevLett.52.2221.

(58) Crawford, O. H. Negative Ions of Polar Molecules. *Molecular Physics* **1971**, *20* (4), 585–591. https://doi.org/10.1080/00268977100100561.

(59) Tripathi, D.; Dutta, A. K. Electron Attachment to DNA Base Pairs: An Interplay of Dipole- and Valence-Bound States. *J. Phys. Chem. A* **2019**, *123* (46), 10131–10138. https://doi.org/10.1021/acs.jpca.9b08974.

(60) Sommerfeld, T. Coupling between Dipole-Bound and Valence States: The Nitromethane Anion. *Phys. Chem. Chem. Phys.* **2002**, *4* (12), 2511–2516. https://doi.org/10.1039/b202143a.

(61) Sommerfeld, T. Intramolecular Electron Transfer from Dipole-Bound to Valence Orbitals: Uracil and 5-Chlorouracil. *J. Phys. Chem. A* **2004**, *108* (42), 9150–9154. https://doi.org/10.1021/jp049082u.

(62) Anstöter, C. S.; Matsika, S. Understanding the Interplay between the Nonvalence and Valence States of the Uracil Anion upon Monohydration. *J. Phys. Chem. A* **2020**, *124* (44), 9237–9243. https://doi.org/10.1021/acs.jpca.0c07407.

(63) Narayanan S J, J.; Tripathi, D.; Dutta, A. K. Doorway Mechanism for Electron Attachment Induced DNA Strand Breaks. *J. Phys. Chem. Lett.* **2021**, *12* (42), 10380–10387. https://doi.org/10.1021/acs.jpclett.1c02735.

(64) Mukherjee, M.; Tripathi, D.; Brehm, M.; Riplinger, C.; Dutta, A. K. Efficient EOM-CC-Based Protocol for the Calculation of Electron Affinity of Solvated Nucleobases: Uracil as a Case Study. *J. Chem. Theory Comput.* **2021**, *17* (1), 105–116. https://doi.org/10.1021/acs.jctc.0c00655.

(65) Mukherjee, M.; Tripathi, D.; Dutta, A. K. Water Mediated Electron Attachment to Nucleobases: Surface-Bound vs Bulk Solvated Electrons. *The Journal of Chemical Physics* **2020**, *153* (4), 044305. https://doi.org/10.1063/5.0010509.

(66) Verma, P.; Ghosh, D.; Dutta, A. K. Electron Attachment to Cytosine: The Role of Water. *J. Phys. Chem. A* **2021**, *125* (22), 4683–4694. https://doi.org/10.1021/acs.jpca.0c10199.

(67) Verma, P.; Narayanan S J, J.; Dutta, A. K. Electron Attachment to DNA: The Protective Role of Amino Acids. *J. Phys. Chem. A* **2023**, *127* (10), 2215–2227. https://doi.org/10.1021/acs.jpca.2c06624.

(68) Narayanan S J, J.; Verma, P.; Adhikary, A.; Kumar Dutta, A. Electron Attachment to the Nucleobase Uracil in Diethylene Glycol: The Signature of a Doorway. *ChemPhysChem* **2024**, *25* (24), e202400581. https://doi.org/10.1002/cphc.202400581.

(69) Gutowski, M.; Dabkowska, I.; Rak, J.; Xu, S.; Nilles, J. M.; Radisic, D.; Bowen Jr, K. H. Barrier-Free Intermolecular Proton Transfer in the Uracil-Glycine Complex Induced by Excess Electron Attachment. *Eur. Phys. J. D* **2002**, *20* (3), 431–439. https://doi.org/10.1140/epjd/e2002-00168-1.

(70) Dąbkowska, I.; Rak, J.; Gutowski, M.; Nilles, J. M.; Stokes, S. T.; Bowen, K. H. Barrier-Free Intermolecular Proton Transfer Induced by Excess Electron Attachment to the Complex of Alanine with Uracil. *The Journal of Chemical Physics* **2004**, *120* (13), 6064–6071. https://doi.org/10.1063/1.1666042.

(71) Dąbkowska, I.; Rak, J.; Gutowski, M.; Nilles, J. M.; Stokes, S. T.; Radisic, D.; Bowen Jr., K. H. Barrier-Free Proton Transfer in Anionic Complex of Thymine with Glycine. *Phys. Chem. Chem. Phys.* **2004**, *6* (17), 4351–4357. https://doi.org/10.1039/B406455K.

(72) Solomun, T.; Skalický, T. The Interaction of a Protein–DNA Surface Complex with Low-Energy Electrons. *Chemical Physics Letters* **2008**, *453* (1–3), 101–104. https://doi.org/10.1016/j.cplett.2007.12.078.

(73) Ptasińska, S.; Li, Z.; Mason, N. J.; Sanche, L. Damage to Amino Acid–Nucleotide Pairs Induced by 1 eV Electrons. *Phys. Chem. Chem. Phys.* **2010**, *12* (32), 9367. https://doi.org/10.1039/b926267a.

(74) Wang, X.; Liao, H.; Liu, W.; Shao, Y.; Zheng, Y.; Sanche, L. DNA Protection against Damages Induced by Low-Energy Electrons: Absolute Cross Sections for Arginine–DNA Complexes. *J. Phys. Chem. Lett.* **2023**, *14* (24), 5674–5680. https://doi.org/10.1021/acs.jpclett.3c01041.

(75) Gu, B.; Smyth, M.; Kohanoff, J. Protection of DNA against Low-Energy Electrons by Amino Acids: A First-Principles Molecular Dynamics Study. *Phys. Chem. Chem. Phys.* **2014**, *16* (44), 24350–24358. https://doi.org/10.1039/C4CP03906H.

(76) Milligan, J. R.; Tran, N. Q.; Ly, A.; Ward, J. F. Peptide Repair of Oxidative DNA Damage. *Biochemistry* **2004**, *43* (17), 5102–5108. https://doi.org/10.1021/bi030232l.

(77) Jena, N. R.; Mishra, P. C.; Suhai, S. Protection Against Radiation-Induced DNA Damage by Amino Acids: A DFT Study. *J. Phys. Chem. B* **2009**, *113* (16), 5633–5644. https://doi.org/10.1021/jp810468m.

(78) Alvarez-Idaboy, J. R.; Galano, A. On the Chemical Repair of DNA Radicals by Glutathione: Hydrogen vs Electron Transfer. *J. Phys. Chem. B* **2012**, *116* (31), 9316–9325. https://doi.org/10.1021/jp303116n.

(79) Marchese, R.; Grandori, R.; Carloni, P.; Raugei, S. On the Zwitterionic Nature of Gas-Phase Peptides and Protein Ions. *PLoS Comput Biol* **2010**, *6* (5), e1000775. https://doi.org/10.1371/journal.pcbi.1000775.

(80) Sarmah, M. P.; Sarma, M. Mechanistic Insights into the Electron Attachment Process to Guanosine in the Presence of Arginine. *Phys. Chem. Chem. Phys.* **2024**, *26* (44), 27955–27963. https://doi.org/10.1039/D4CP02558J.

(81) Pracht, P.; Bohle, F.; Grimme, S. Automated Exploration of the Low-Energy Chemical Space with Fast Quantum Chemical Methods. *Phys. Chem. Chem. Phys.* **2020**, *22* (14), 7169–7192. https://doi.org/10.1039/C9CP06869D.

(82) Sarmah, M. P.; Medhi, B.; Sarma, M. Impact of the Electron Attachment to the Alanyl Glycine and Glycyl Alanine Conformers. *The Journal of Chemical Physics* **2026**, *164* (5), 054307. https://doi.org/10.1063/5.0297297.

(83) Dutta, A. K.; Saitow, M.; Demoulin, B.; Neese, F.; Izsák, R. A Domain-Based Local Pair Natural Orbital Implementation of the Equation of Motion Coupled Cluster Method for Electron Attached States. *The Journal of Chemical Physics* **2019**, *150* (16), 164123. https://doi.org/10.1063/1.5089637.

(84) Sawicka, A.; Anusiewicz, I.; Skurski, P.; Simons, J. Dipole-bound Anions Supported by Charge–Transfer Interaction: Anionic States of $H_n F_{3-n} N \rightarrow BH_3$ and $H_3 N \rightarrow BH_n F_{3-n}$ ($n$ = 0, 1, 2, 3). *Int J of Quantum Chemistry* **2003**, *92* (4), 367–375. https://doi.org/10.1002/qua.10517.

(85) Phillips, J. C.; Braun, R.; Wang, W.; Gumbart, J.; Tajkhorshid, E.; Villa, E.; Chipot, C.; Skeel, R. D.; Kalé, L.; Schulten, K. Scalable Molecular Dynamics with NAMD. *J Comput Chem* **2005**, *26* (16), 1781–1802. https://doi.org/10.1002/jcc.20289.

(86) Neese, F.; Wennmohs, F.; Becker, U.; Riplinger, C. The ORCA Quantum Chemistry Program Package. *The Journal of Chemical Physics* **2020**, *152* (22), 224108. https://doi.org/10.1063/5.0004608.

(87) Neese, F. Software Update: The ORCA Program System—Version 5.0. *WIREs Comput Mol Sci* **2022**, *12* (5), e1606. https://doi.org/10.1002/wcms.1606.

(88) Svozil, D.; Frigato, T.; Havlas, Z.; Jungwirth, P. Ab Initio Electronic Structure of Thymine Anions. *Phys. Chem. Chem. Phys.* **2005**, *7* (5), 840. https://doi.org/10.1039/b415007d.

(89) Tripathi, D.; Dutta, A. K. Bound Anionic States of DNA and RNA Nucleobases: An EOM-CCSD Investigation. *Int J of Quantum Chemistry* **2019**, *119* (9), e25875. https://doi.org/10.1002/qua.25875.

(90) Tyburski, R.; Liu, T.; Glover, S. D.; Hammarström, L. Proton-Coupled Electron Transfer Guidelines, Fair and Square. *J. Am. Chem. Soc.* **2021**, *143* (2), 560–576. https://doi.org/10.1021/jacs.0c09106.

(91) Köuppel, H.; Domcke, W.; Cederbaum, L. S. Multimode Molecular Dynamics Beyond the Born-Oppenheimer Approximation. In *Advances in Chemical Physics*; Prigogine, I., Rice, S. A., Eds.; Wiley, 1984; Vol. 57, pp 59–246. https://doi.org/10.1002/9780470142813.ch2.

(92) Marcus, R. A. On the Theory of Oxidation-Reduction Reactions Involving Electron Transfer. I. *The Journal of Chemical Physics* **1956**, *24* (5), 966–978. https://doi.org/10.1063/1.1742723.

(93) Mukherjee, M.; Tripathi, D.; Dutta, A. K. Water Mediated Electron Attachment to Nucleobases: Surface-Bound vs Bulk Solvated Electrons. *The Journal of Chemical Physics* **2020**, *153* (4), 044305. https://doi.org/10.1063/5.0010509.

(94) Yu, D.; Rauk, A.; Armstrong, D. A. Radicals and Ions of Glycine: An Ab Initio Study of the Structures and Gas-Phase Thermochemistry. *J. Am. Chem. Soc.* **1995**, *117* (6), 1789–1796. https://doi.org/10.1021/ja00111a017.

(95) Nacsa, A. B.; Czakó, G. Benchmark *Ab Initio* Proton Affinity of Glycine. *Phys. Chem. Chem. Phys.* **2021**, *23* (16), 9663–9671. https://doi.org/10.1039/D1CP00376C.

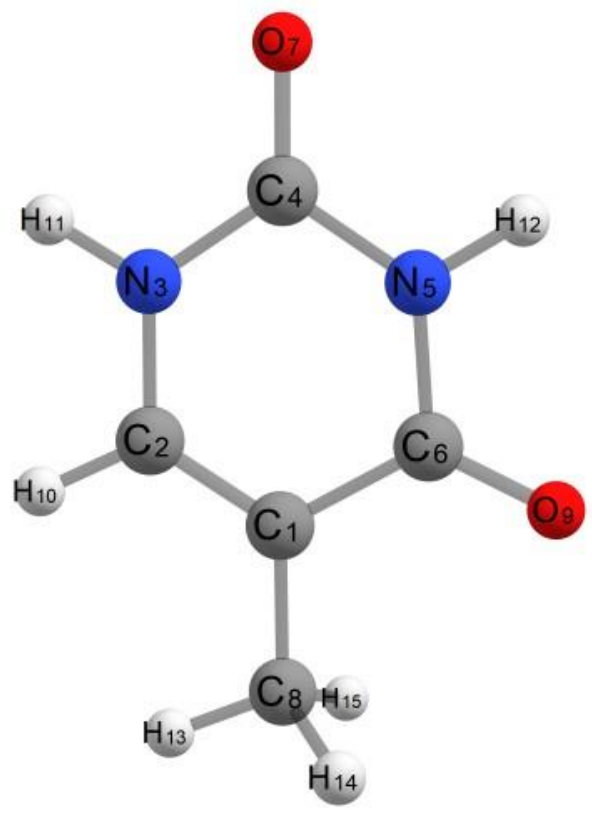

**Figure 1. Optimized structure of thymine at neutral geometry.**

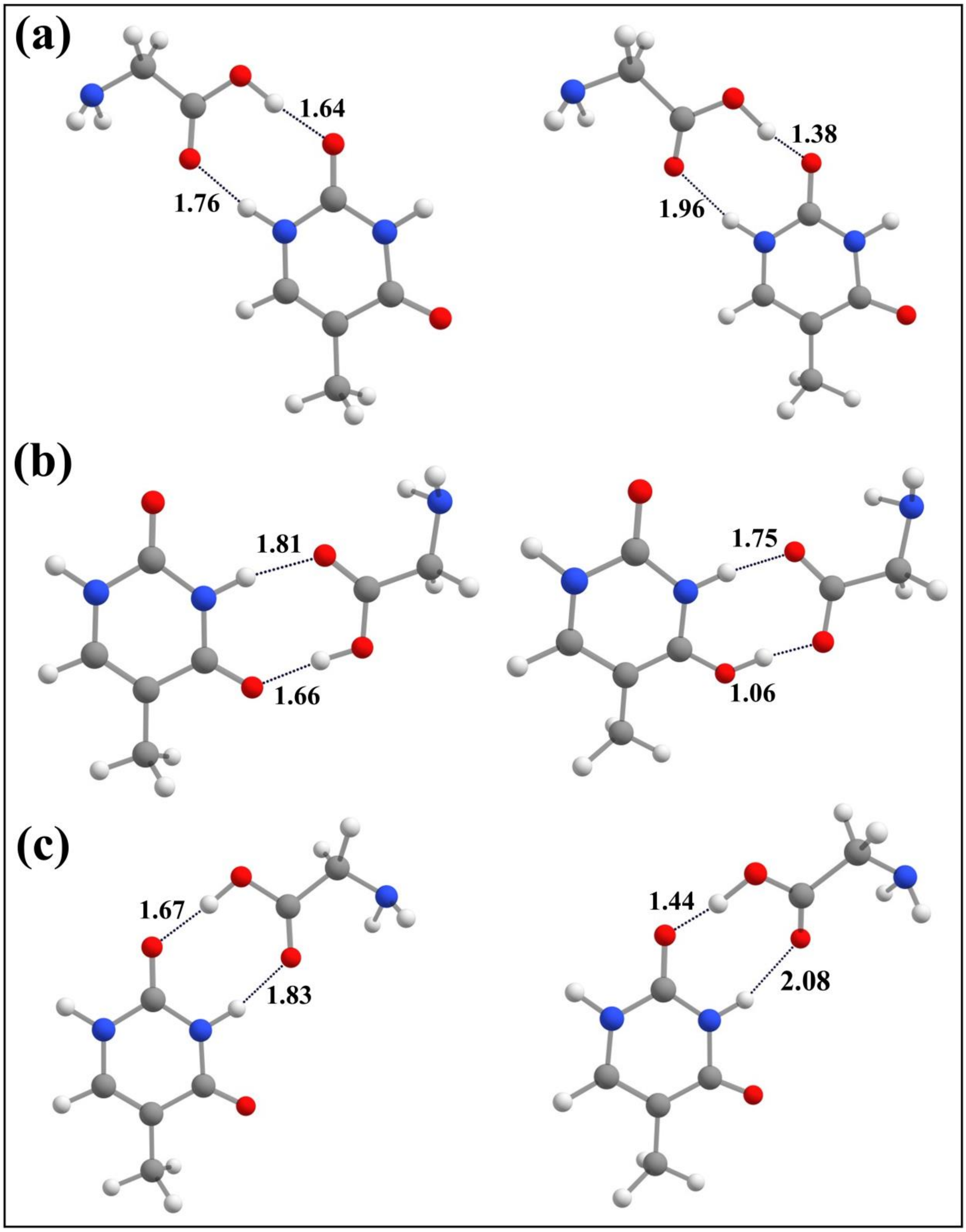

**Figure 2. Optimized structures for (a) Thy-NGly1 (b)Thy-NGly2 and (c) Thy-NGly3 at neutral (left) and anionic (right) geometry. The numbers indicate hydrogen-bond distances in Å.**

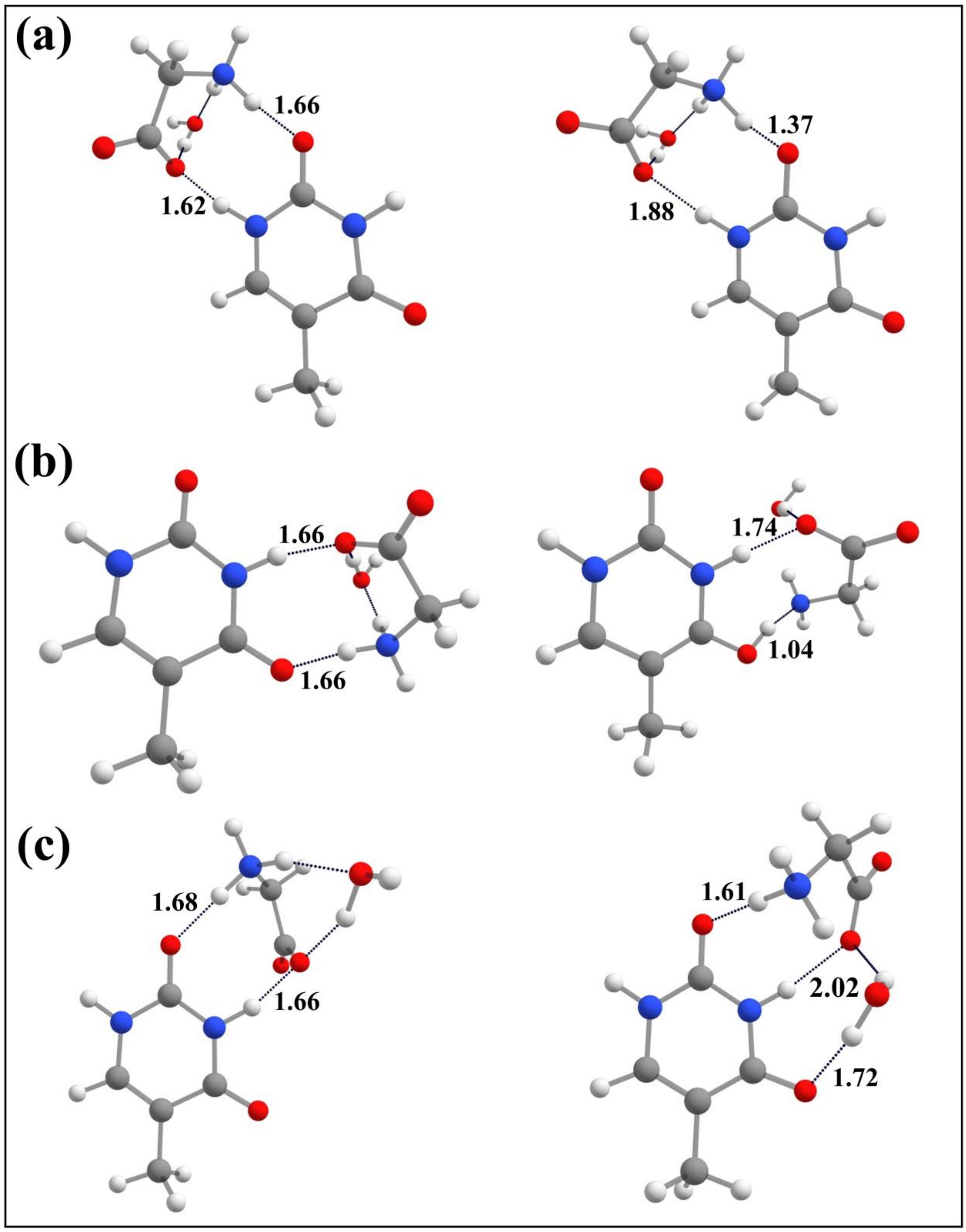

**Figure 3. Optimized structures for (a) Thy-ZGly1 (b)Thy-ZGly2 and (c) Thy-ZGly3 at neutral (left) and anionic (right) geometry. The numbers indicate hydrogen-bond distances in Å.**

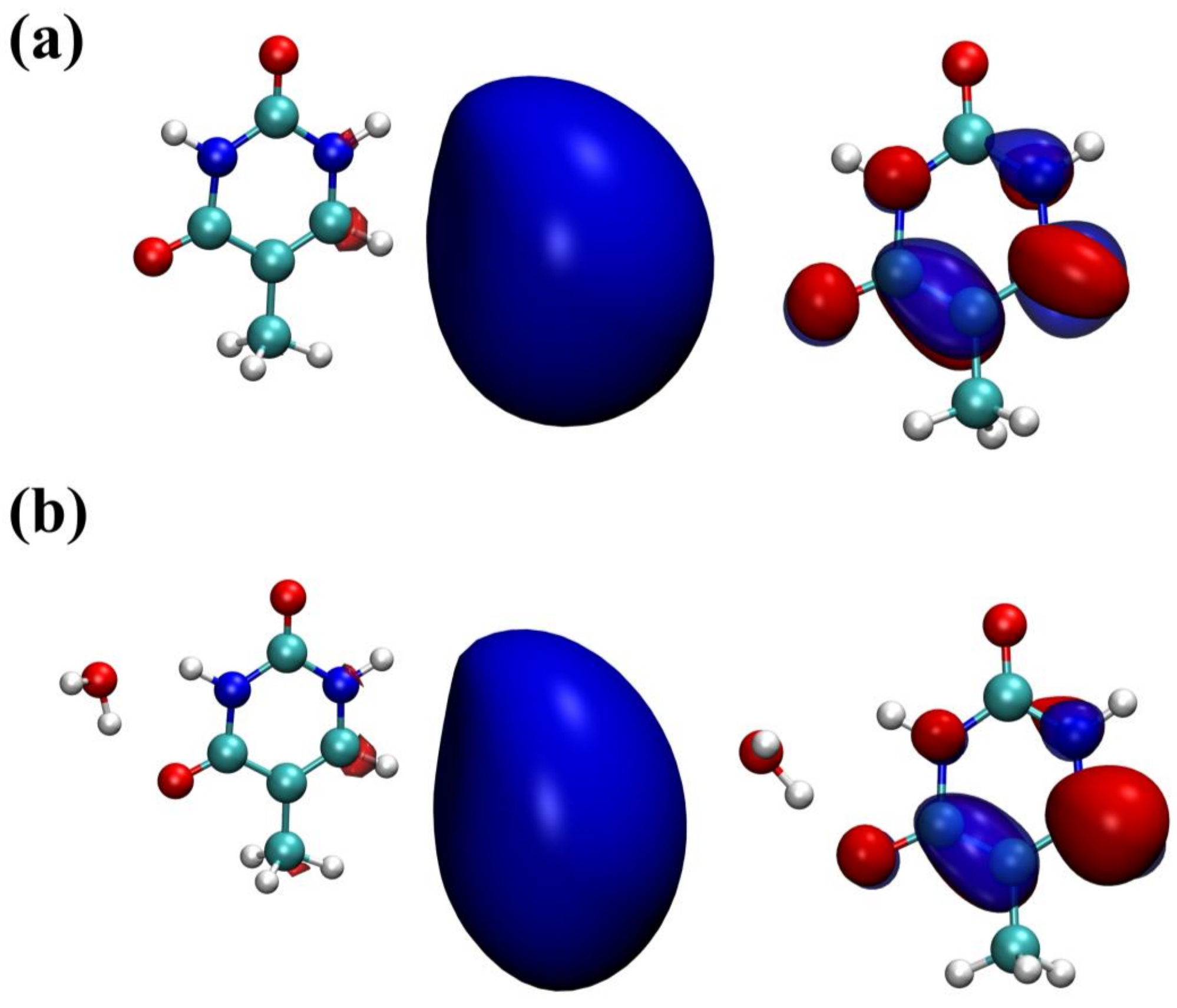

**Figure 4. EA-EOM-DLPNO-CCSD natural orbitals corresponding to the dipole bound anions (left) and valence bound anions (right) for (a) isolated-thymine and (b) thymine-water.**

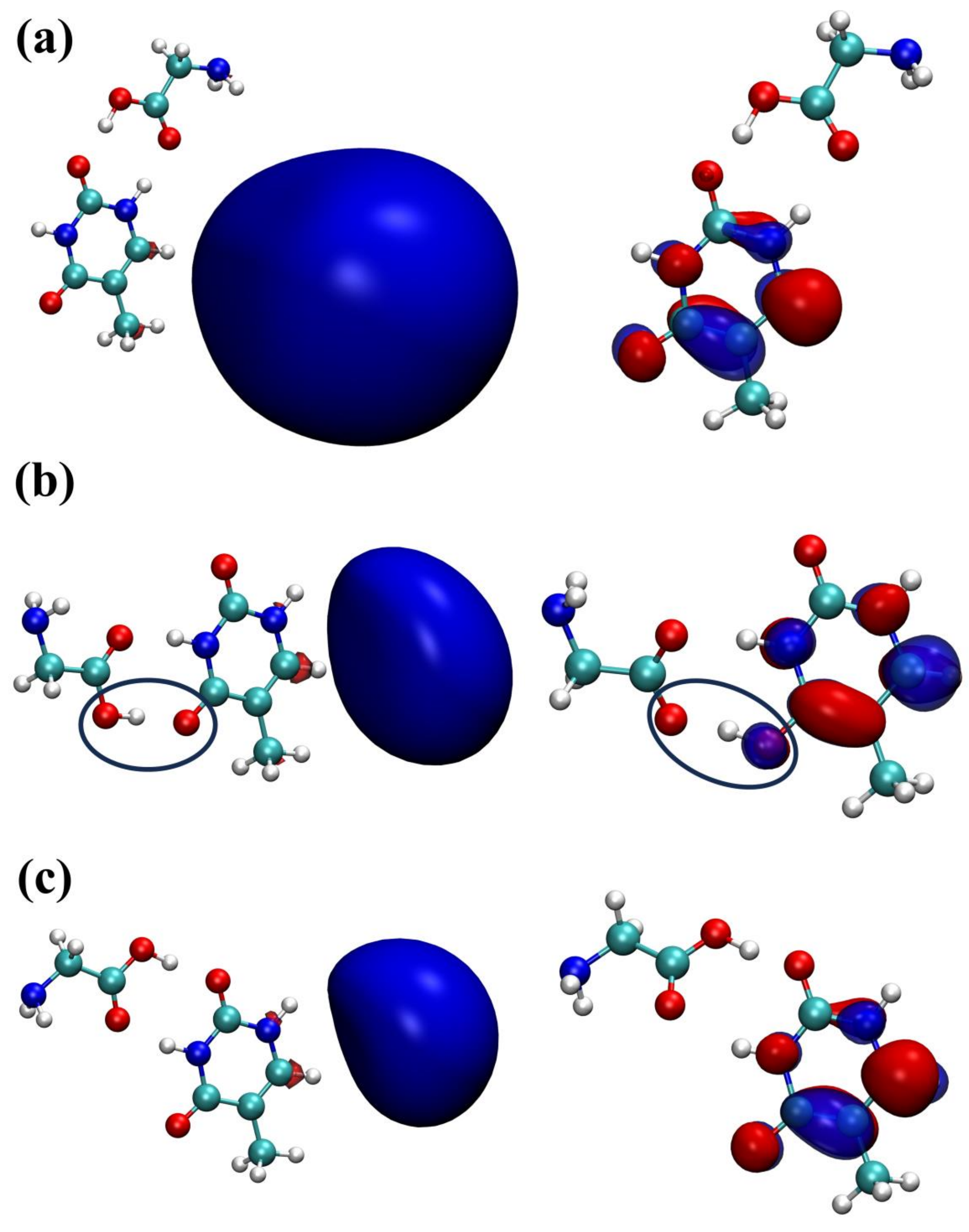

**Figure 5. EA-EOM-DLPNO-CCSD natural orbitals corresponding to the dipole bound anions (left) and valence bound anions (right) for (a)Thy-NGly1, (b) Thy-NGly2, (c)Thy-NGly3.**

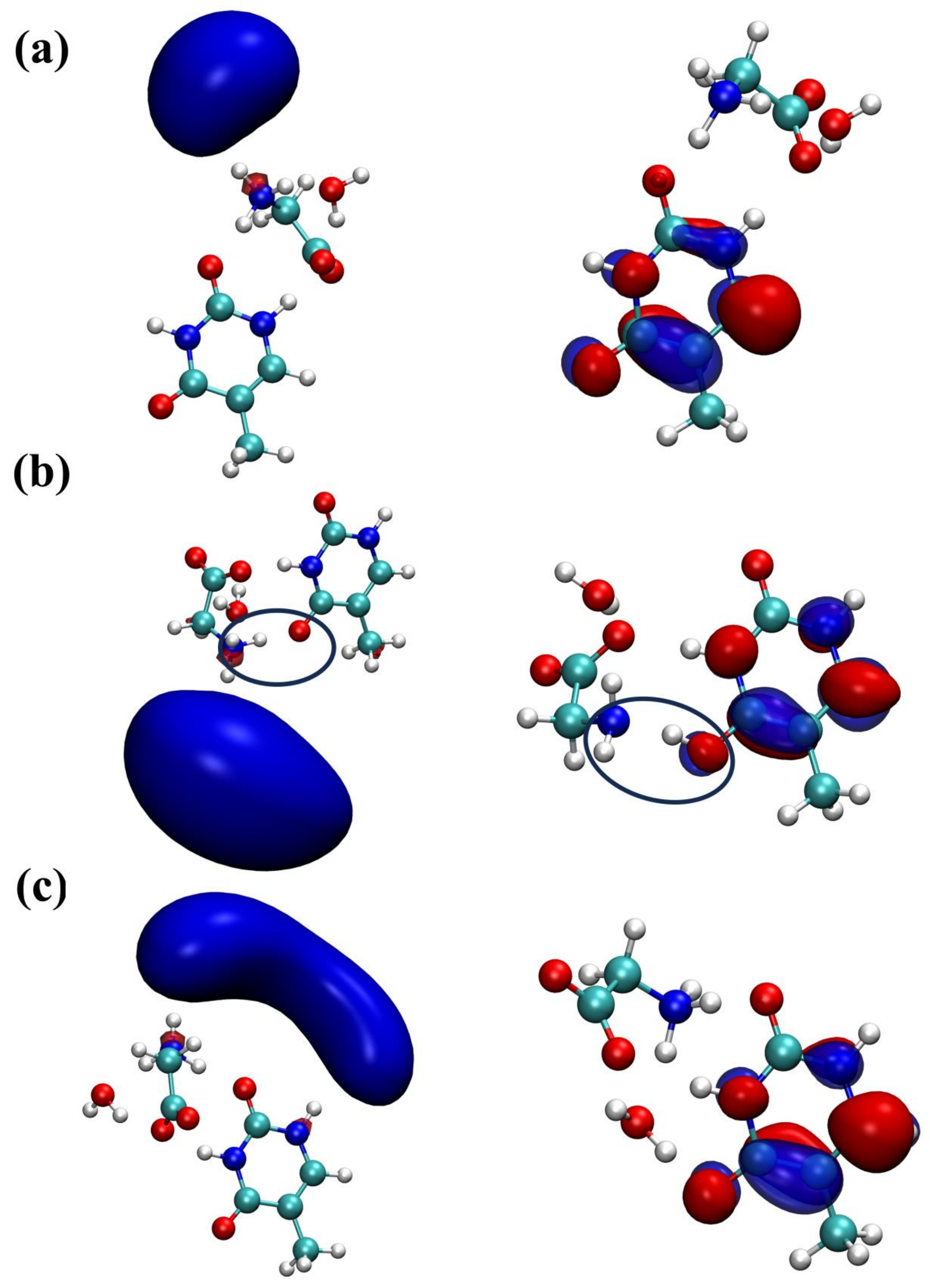

**Figure 6. EA-EOM-DLPNO-CCSD natural orbitals corresponding to the dipole bound anions (left) and valence bound anions (right) for (a)Thy-ZGly1, (b) Thy-ZGly2, (c)Thy-ZGly3.**

***Table 1. Dipole moment (in Debye) and VEA, VDE and AEA (all values are in meV) of Isolated Thymine, Thymine-Water and Thymine-Glycine Complexes.***

| Complex | Dipole moment | VEA | VDE | AEA |
|---|---|---|---|---|
| Isolated thymine | 4.88 | 48 | 570 | -231 |
| Thymine-Water | 4.95 | 40 | 937 | -4 |
| Thy-$NGly_1$ | 3.18 | 4 | 862 | -90 |
| Thy-$NGly_2$ | 6.05 | 65 | 1883 | 300 |
| Thy-$NGly_3$ | 6.15 | 74 | 865 | -103 |
| Thy-$ZGly_1$ | 6.04 | 127 | 890 | -109 |
| Thy-$ZGly_2$ | 8.70 | 77 | 2125 | 563 |
| Thy-$ZGly_3$ | 10.30 | 132 | 1456 | 315 |

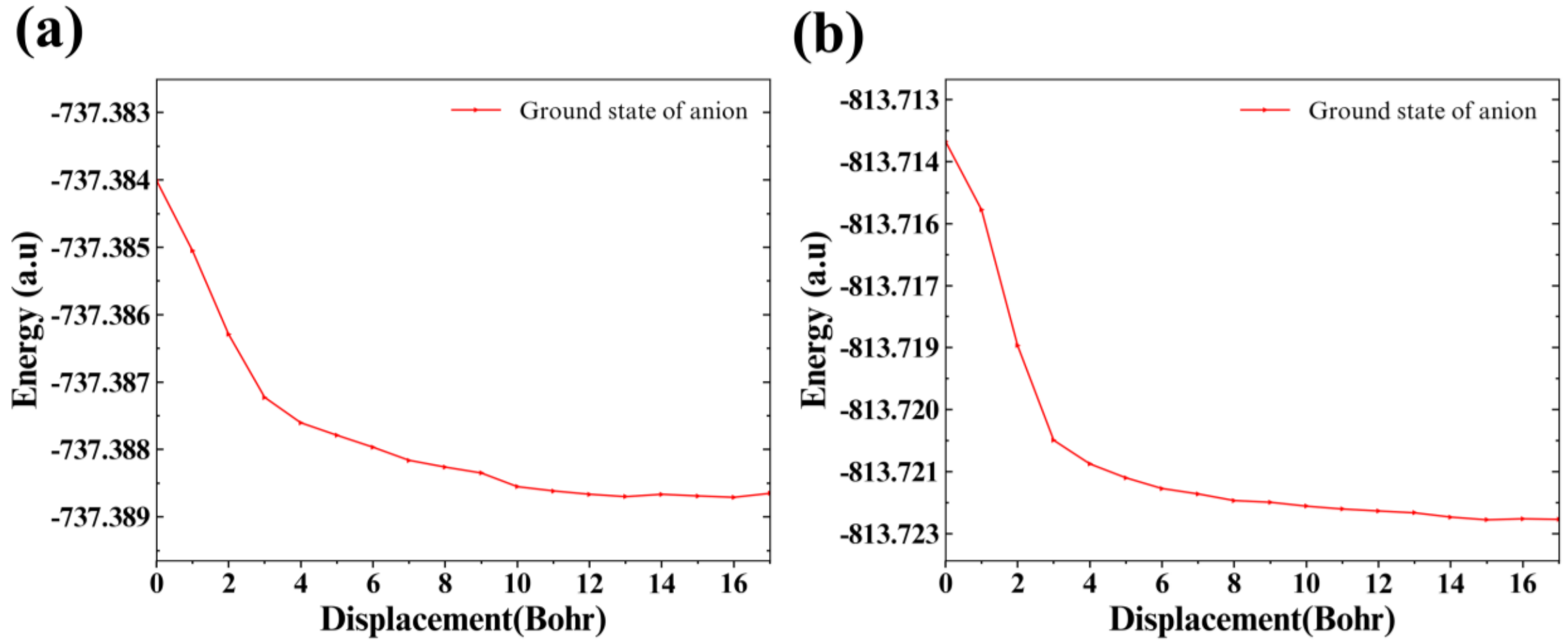

**Figure 7. MEP for the proton transfer reaction in (a)Thy-NGly2 and (b)Thy-ZGly2.**

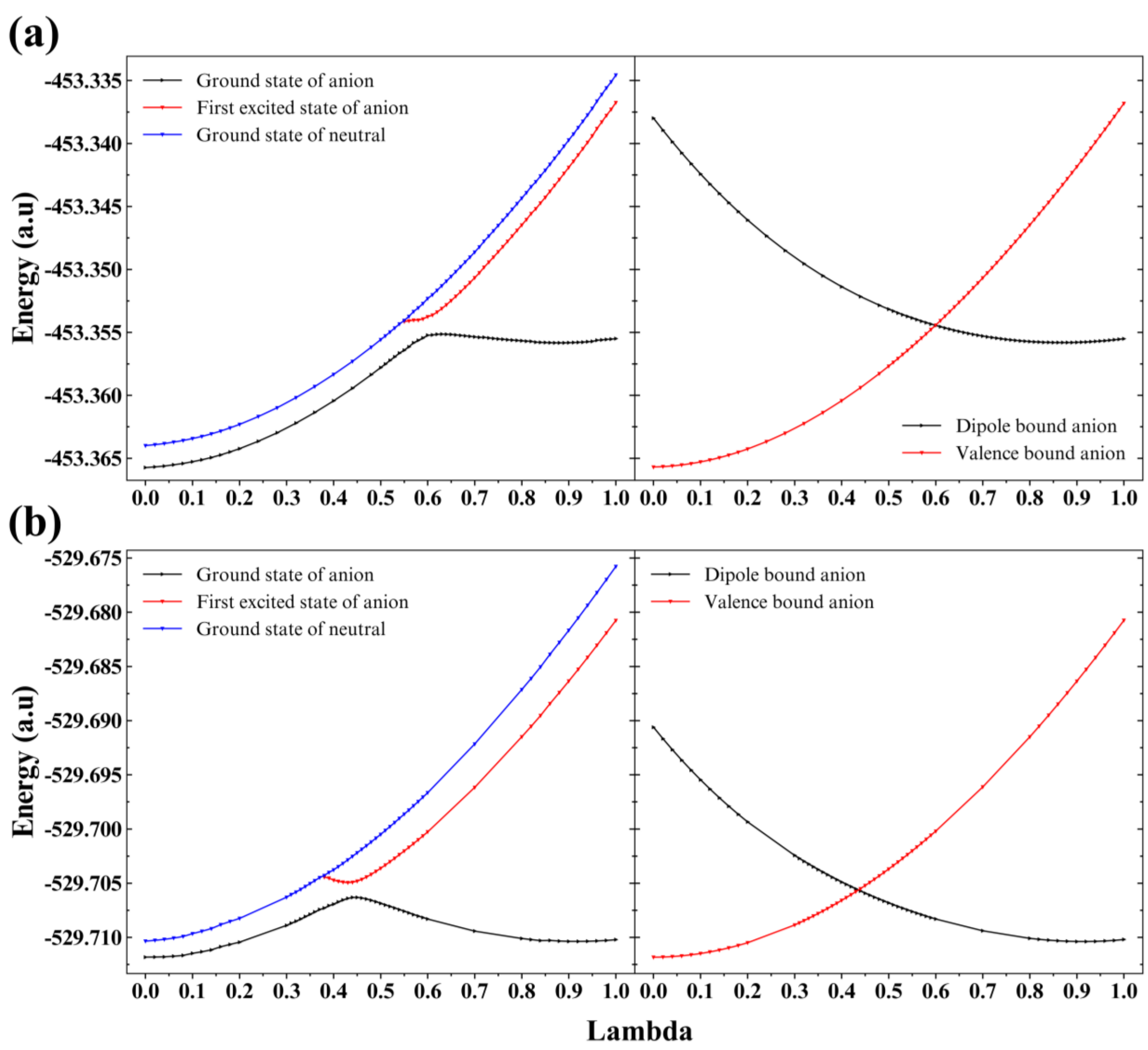

**Figure 8. Adiabatic (left) and diabatic (right) PECs for (a) isolated thymine and (b) thymine-water complex.**

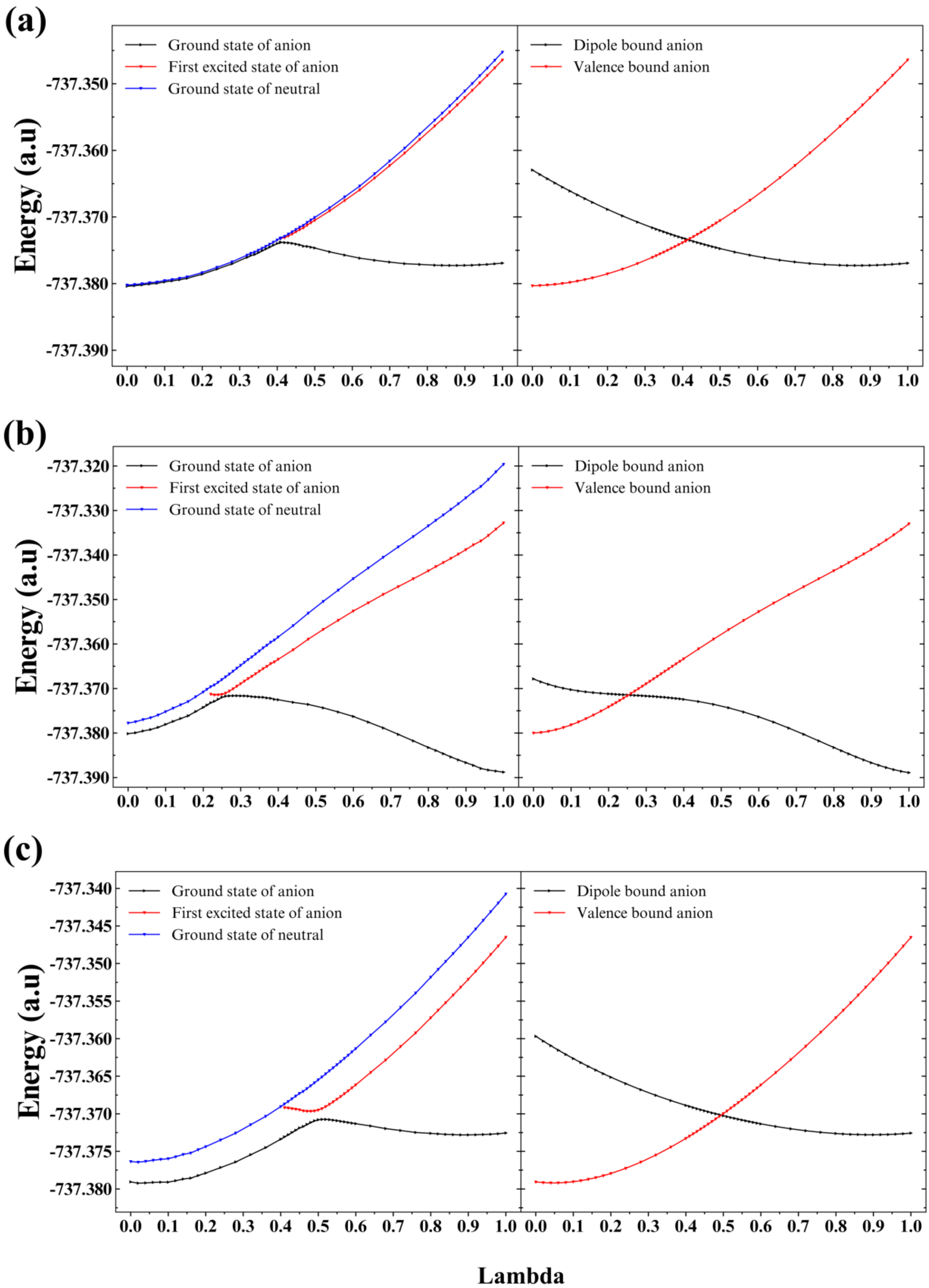

**Figure 9. Adiabatic (left) and diabatic (right) PECs for (a) Thy-NGly1, (b) Thy-NGly2, (c) Thy-NGly3.**

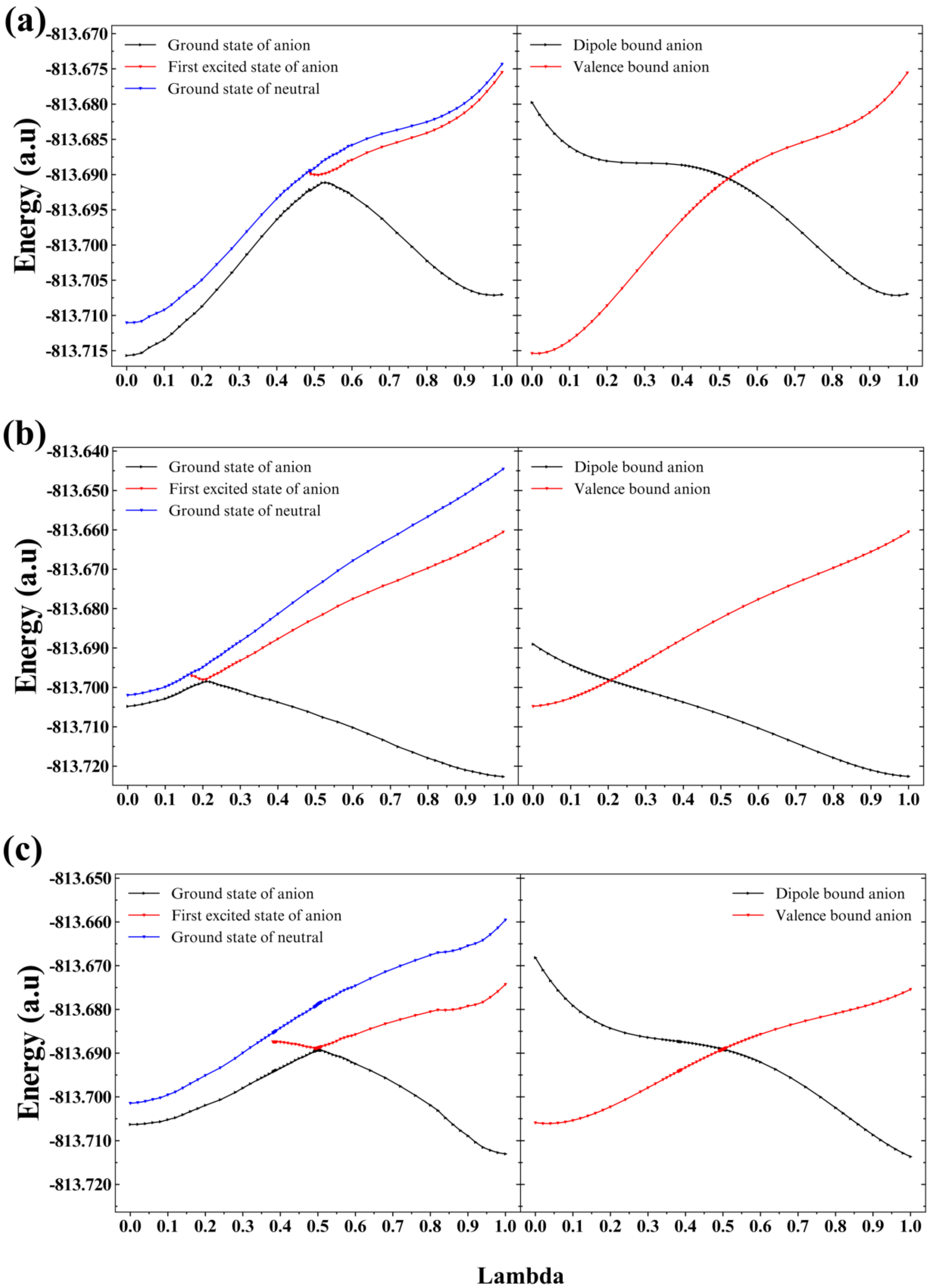

**Figure 10. Adiabatic (left) and diabatic (right) PECs for (a) Thy-ZGly1, (b) Thy-ZGly2, (c) Thy-ZGly3.**

***Table 2. Coupling Constant (meV) and Rate Constant ($s^{-1}$) of Electron Transfer from the Dipole Bound to Valence Bound State at the EA-EOM-DLPNO-CCSD/aug-cc-pVTZ(+5s5p4d) Level of Theory***

| Complex | Coupling constant | Rate constant |
|---|---|---|
| Thymine | 21.68 | $1.0\times10^{8}$ |
| Thymine-Water | 18.87 | $1.8\times10^{10}$ |
| Thy-NGly1 | 9.50 | $8.1\times10^{9}$ |
| Thy-NGly2 | 11.35 | $4.0\times10^{11}$ |
| Thy-NGly3 | 18.14 | $4.2\times10^{9}$ |
| Thy-ZGly1 | 15.85 | $1.5\times10^{7}$ |
| Thy-ZGly2 | 7.14 | $1.2\times10^{11}$ |
| Thy-ZGly3 | 4.93 | $5.1\times10^{7}$ |

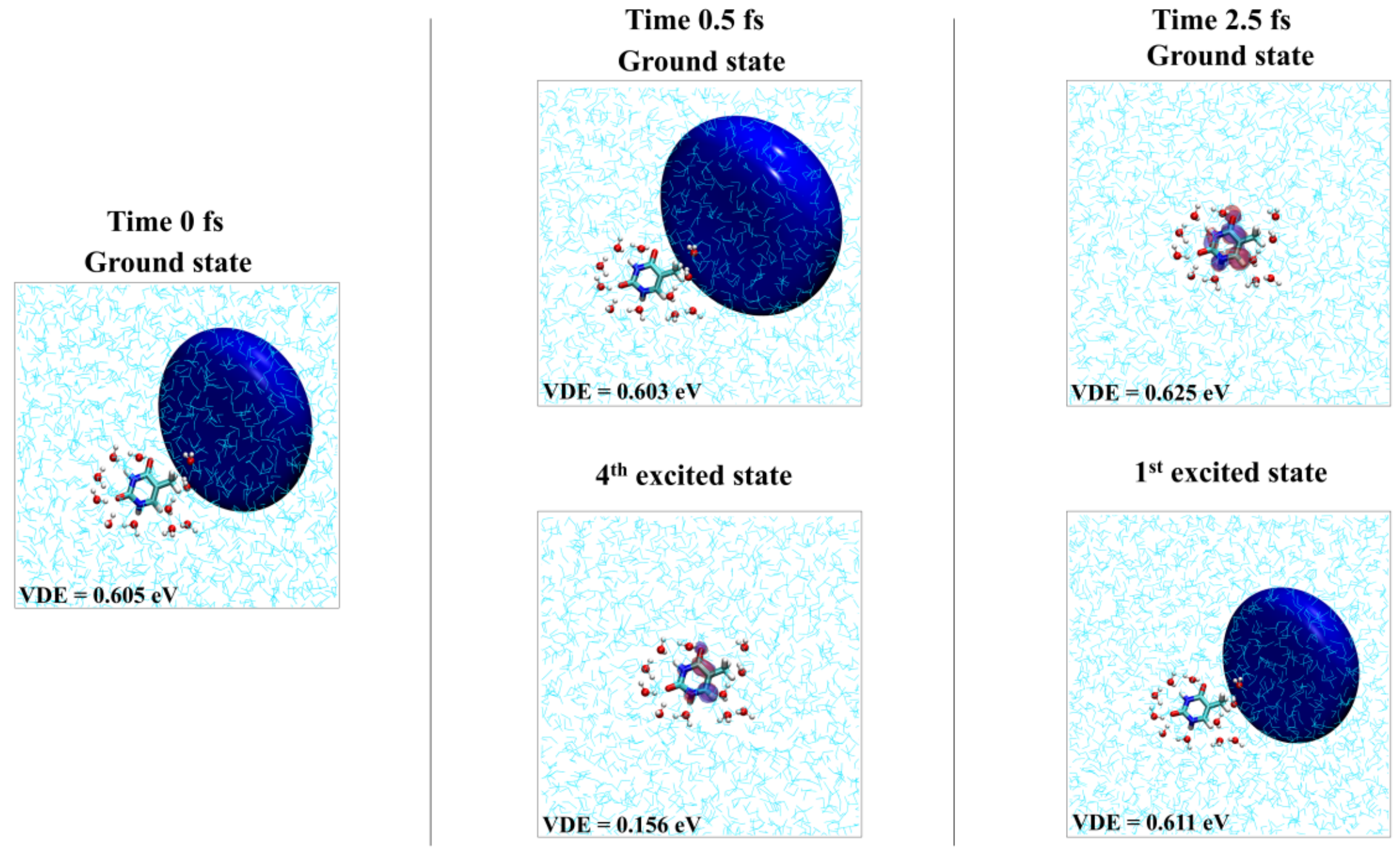

**Figure 11. EA-EOM-DLPNO-CCSD dominant transition orbitals depicting the time evolution of anionic state for thymine solvated in water.**

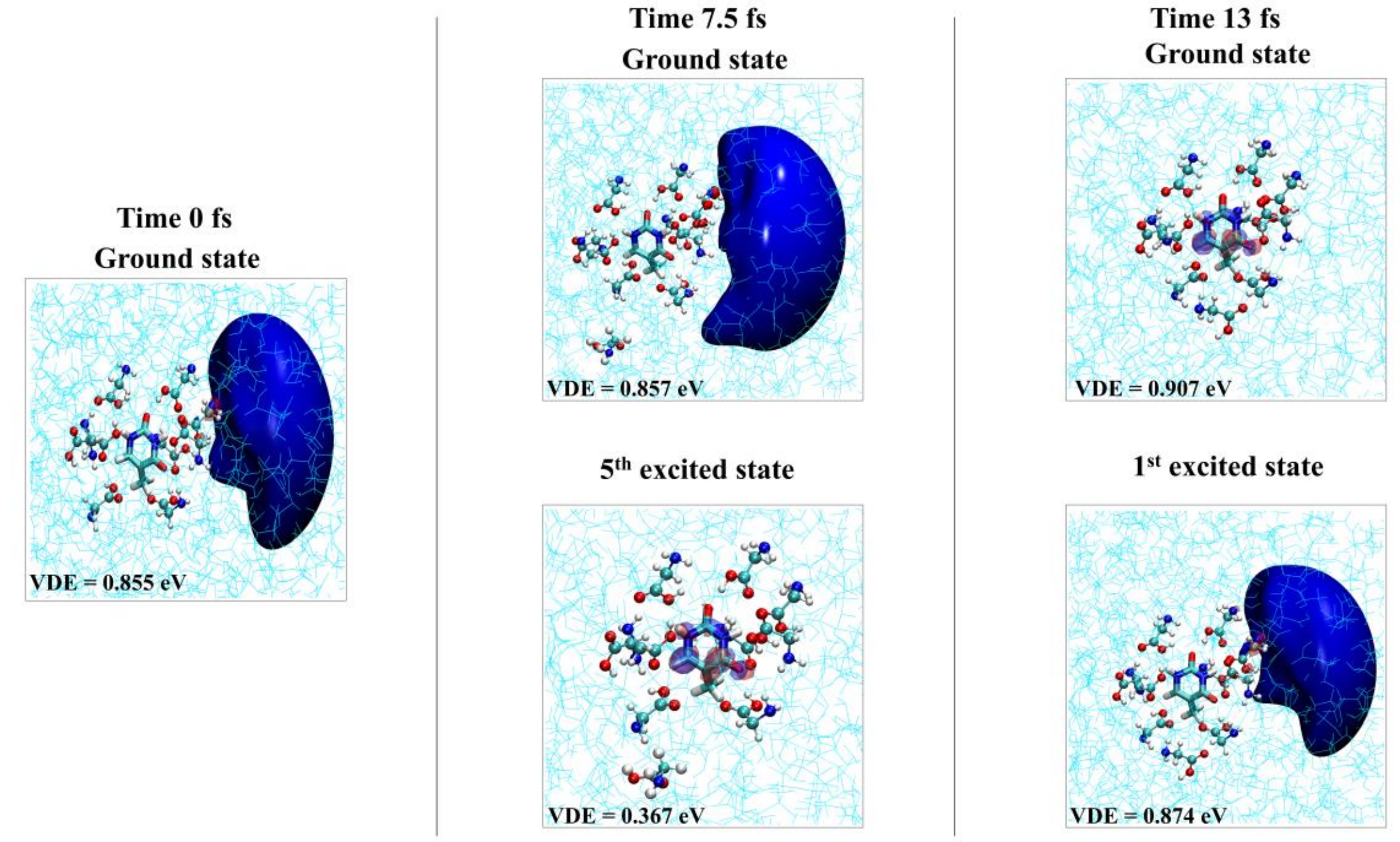

**Figure 12. EA-EOM-DLPNO-CCSD dominant transition orbitals depicting the time evolution of anionic state for thymine solvated in native glycine.**

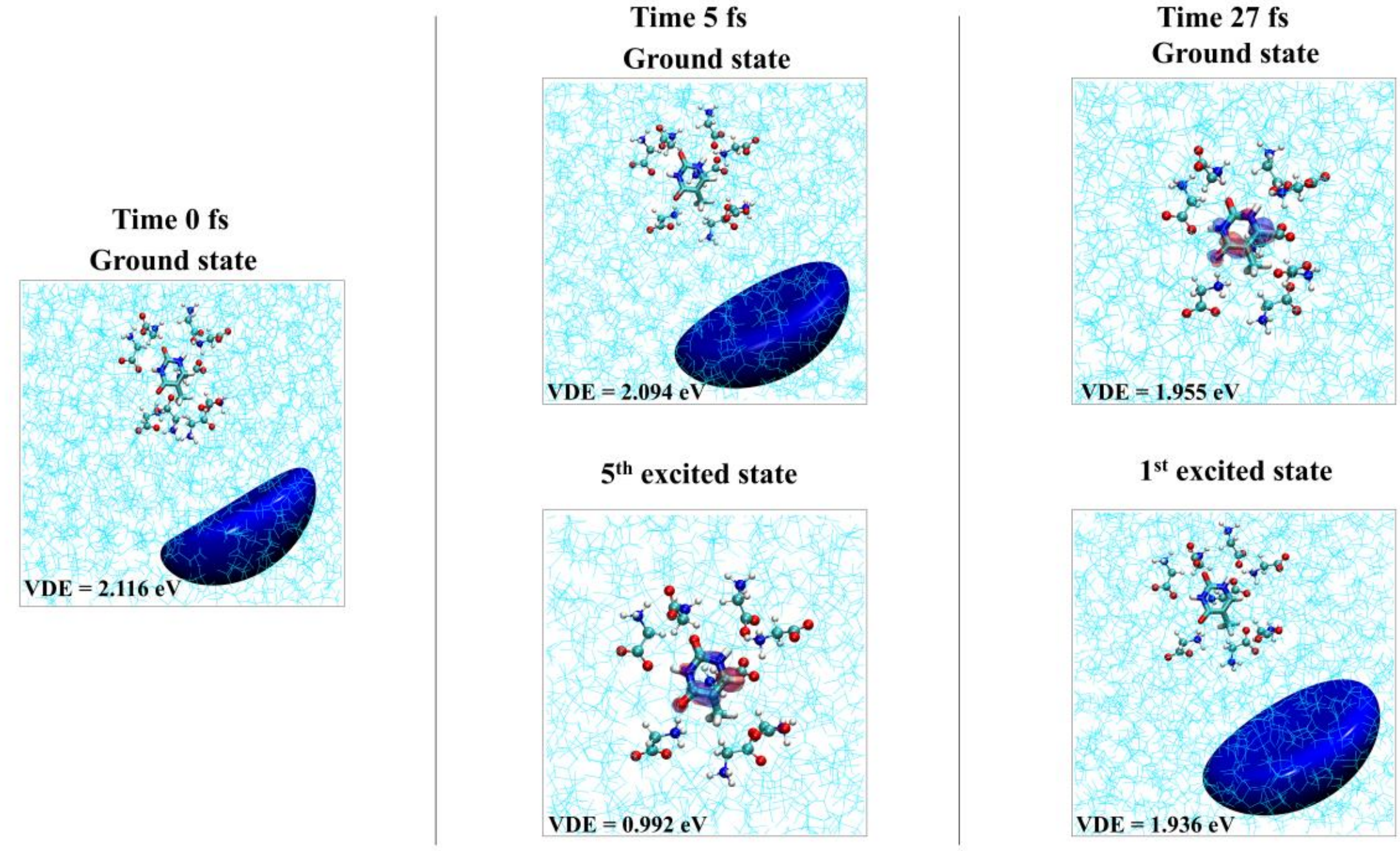

**Figure 13. EA-EOM-DLPNO-CCSD dominant transition orbitals depicting the time evolution of anionic state for thymine solvated in zwitterionic glycine.**

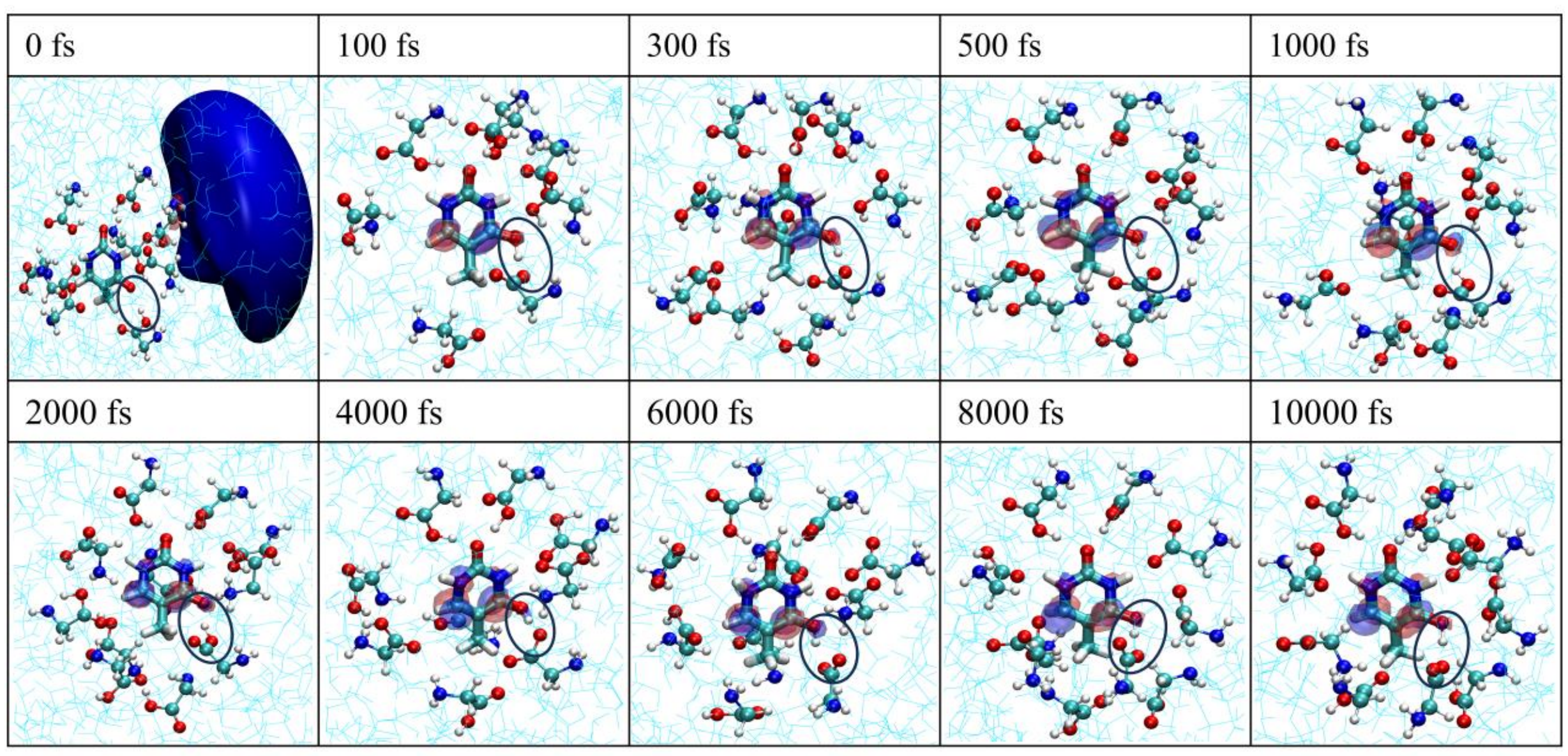

**Figure 14. Evolution of the ground anionic state of thymine in a bulk native glycine environment, illustrating the proton transfer from native glycine to thymine in the first QM/MM trajectory**

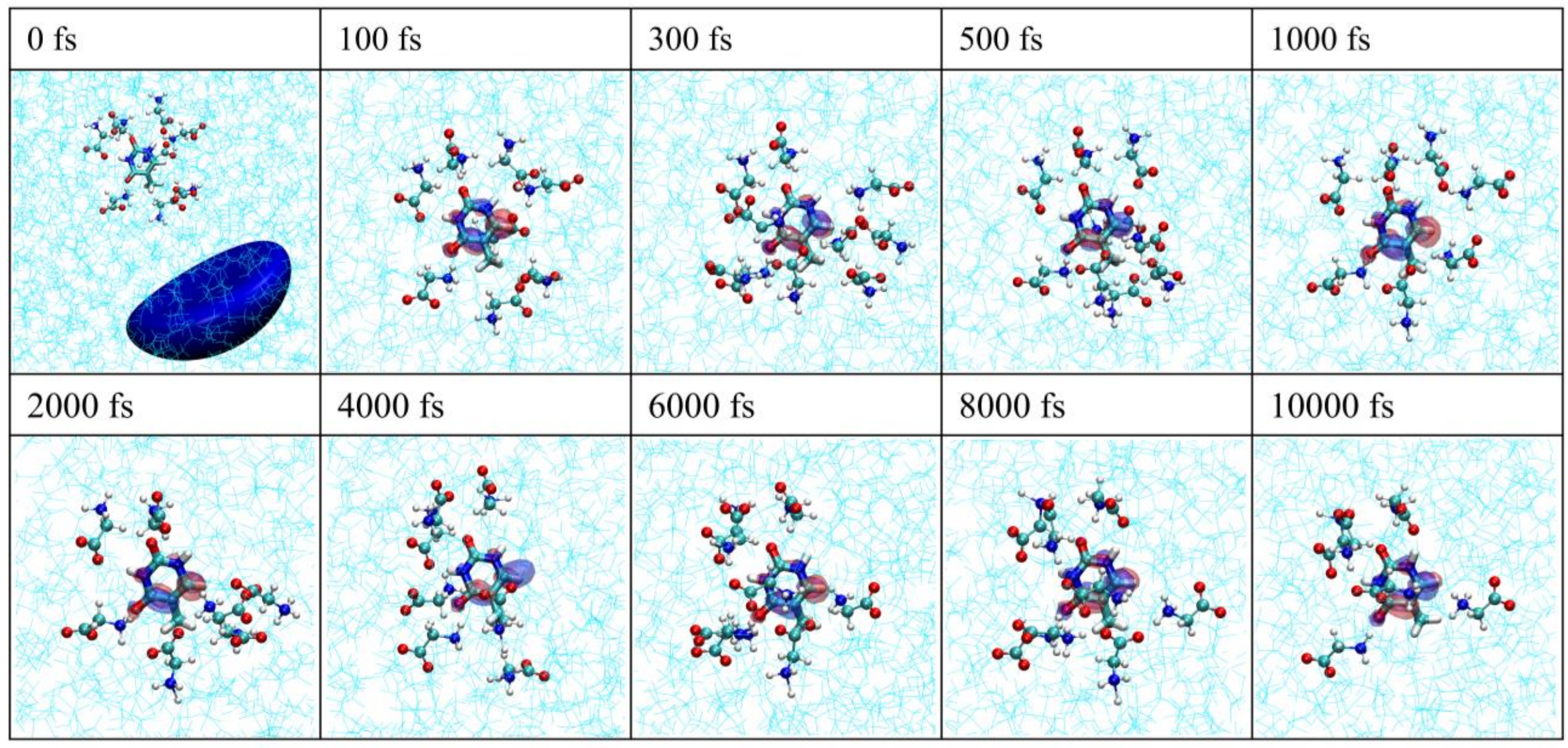

**Figure 15. Evolution of the ground anionic state of thymine in a bulk zwitterionic glycine environment in the first QM/MM trajectory.**

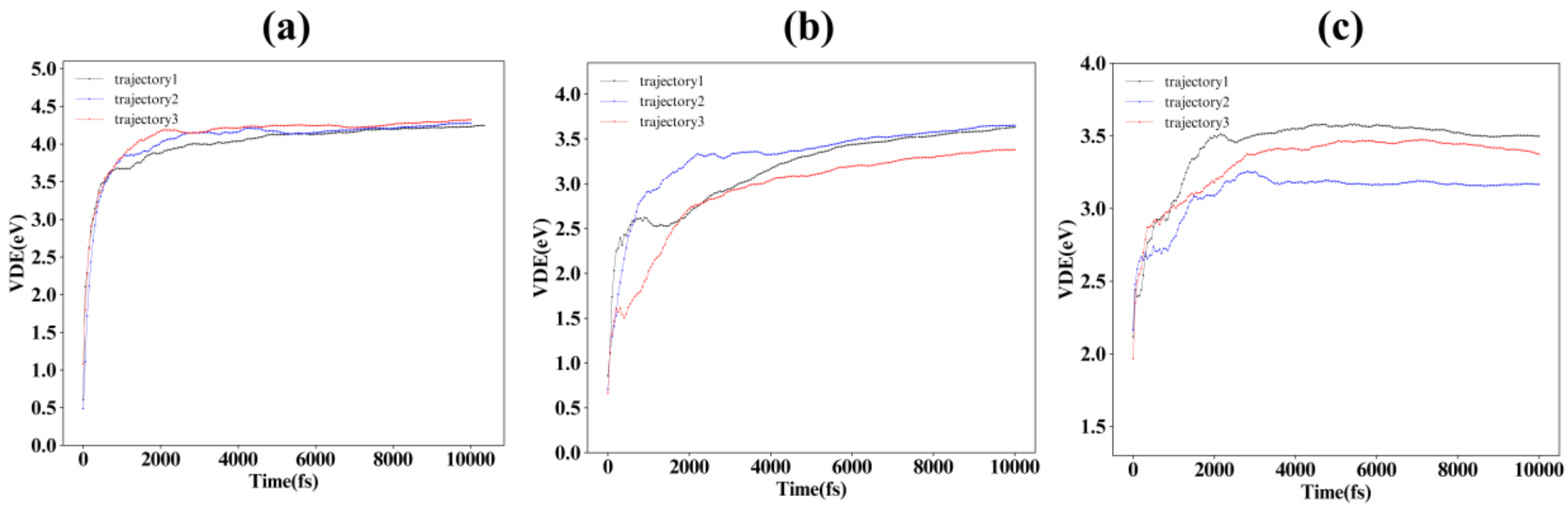

**Figure 16. Instantaneous average of the VDE for the ground anionic state of (a) thymine-water (b) thymine-native glycine and (c) thymine-zwitterionic glycine in the QM/MM trajectories.**